\documentclass[%
 reprint,
superscriptaddress,
 amsmath,amssymb,
 aps,
floatfix,
]{revtex4-2}

\usepackage{graphicx}
\usepackage{dcolumn}
\usepackage{bm}
\usepackage{xcolor}

\begin{document}

\title{Absence of universal backscattering immunity in reciprocal photonic waveguides}

\author{Yecheng Hu}
\author{Ming Gao}
\author{Wei Xie}
\author{Yinxiang Xiong}
\author{Canqiu Dong}
\affiliation{%
Electrical and Computer Engineering, University of Sydney,
Sydney, NSW 2006, Australia
}

\author{Zhuoyuan Lu}
\author{Yuerui Lu}
\affiliation{%
School of Engineering, College of Engineering, Computing and Cybernetics,
The Australian National University, Canberra, ACT 2601, Australia
}
\affiliation{%
ARC Centre for Quantum Computation and Communication Technology,
The Australian National University, Canberra, ACT 2601, Australia
}

\author{Sid Assawaworrarit}
\affiliation{%
Electrical and Computer Engineering, University of Sydney,
Sydney, NSW 2006, Australia
}

\author{Alex Y. Song}
\email{alex.song@sydney.edu.au}
\affiliation{%
Electrical and Computer Engineering, University of Sydney,
Sydney, NSW 2006, Australia
}

\begin{abstract}

Suppressing parasitic back-reflection caused by imperfections, such as random defects, material inhomogeneities, localized inclusions, and fabrication-induced geometric deviations, has long been a central objective in optical-waveguide engineering. In this Letter, we theoretically establish that universal immunity to backscattering from arbitrary disorder is fundamentally impossible in reciprocal optical waveguides, irrespective of whether the underlying structure is topologically trivial or nontrivial.We establish this result through two complementary constructions of physically admissible reciprocal perturbations. The first considers weak permittivity perturbations of finite spatial extent within the first Born approximation, whereas the second considers optically small dielectric perturbations of finite contrast. In each construction, an admissible realization exists that produces a nonzero leading-order backward-scattering amplitude. We trace the fundamental origin of this limitation to the absence in photons of the intrinsic spin-$\tfrac{1}{2}$ and fermionic time-reversal structure underlying Kramers protection in electronic systems. Consequently, any mechanism that enforces photonic backscattering suppression throughout a prescribed class of disorder must instead be engineered into the underlying electromagnetic structure and its constitutive response, which can itself be modified by generic reciprocal perturbations. Numerical examples in representative reciprocal topological waveguides further illustrate these limitations for random defects and sharp bends. We further identify several constructive strategies for suppressing back-reflection within certain prescribed classes of disorder.

\end{abstract}

\maketitle


\section{\label{sec:level1}Introduction}
Suppressing parasitic back-reflection induced by imperfections has been a longstanding objective in optical-waveguide engineering. This requirement becomes increasingly stringent as nanophotonic circuits scale in complexity and integration density, where fabrication-induced reflections from multiple waveguides and components can accumulate into appreciable circuit-level effects~\cite{Morichetti2010,Thomson2016,Bogaerts2018}. In quantum-photonic systems, stringent suppression of parasitic scattering and back-reflection is likewise required to preserve high-fidelity interference, correlations, entanglement, and controlled transport of nonclassical states~\cite{Wang2020,BlancoRedondo2020,Yan2021,Ozawa2019,Mehrabad2023}.

Genuinely backscattering-free waveguiding can in principle be realized in nonreciprocal photonic systems through either magnetic bias~\cite{haldane2008possible,wang2008reflection,wang2009observation} or spatiotemporal modulation~\cite{fang2012effective,sounas2017nonreciprocal,yu2009isolation,lira2012electrically}. The vast majority of optical waveguides, however, are formed from static, non-magnetically biased media, including vacuum, dielectrics, semiconductors, metals, and plasmonic materials, and therefore belong to the broad class of reciprocal systems~\cite{potton2004reciprocity,jalas2013isolator,caloz2018nonreciprocity}. Reciprocity is one of the most fundamental symmetry principles in photonics~\cite{potton2004reciprocity,sounas2017nonreciprocal}. Implementing strong nonreciprocity in such optical platforms remains extremely challenging at optical frequencies, owing to the material, loss, and biasing constraints of magneto-optical approaches and the high-speed modulation and driving complexity required by spatiotemporal schemes. This therefore raises the central question of whether it is possible to design a reciprocal optical waveguide whose transport remains robust against arbitrary localized reciprocal imperfections and defects, ideally allowing light to propagate without back-reflection.

In recent years, topological photonics has emerged as a prominent route toward robust waveguiding in reciprocal systems. In particular, time-reversal-invariant photonic topological phases can be realized without breaking electromagnetic reciprocity, including photonic analogues of quantum spin-Hall and valley-Hall systems~\cite{khanikaev2013photonic,wu2015scheme,ma2016allsi,chen2017valley,noh2018observation,shalaev2019robust}. Their suppression of backscattering, however, relies on additional constraints beyond reciprocity itself, including protecting symmetries and the associated pseudospin selection rules in photonic quantum spin-Hall systems, and momentum-space separation that suppresses intervalley coupling in photonic valley-Hall systems. Recent theoretical and experimental studies increasingly point to limitations in the backscattering protection of those reciprocal photonic topological waveguides~\cite{orazbayev2019quantitative,saba2020nature,shah2021tunneling,rosiek2023observation}.

In this Letter, we establish that universal immunity to disorder-induced backscattering is fundamentally impossible in reciprocal optical waveguides, irrespective of whether the underlying structure is topologically trivial or nontrivial. In the main text, we establish this result for reciprocal perturbations that modify only the permittivity tensor, with the extension to perturbations of the full constitutive matrix given in the Supplementary Material. We consider two experimentally relevant limits: a finite-extent, weak permittivity perturbation treated within the first Born approximation, and an electrically small, nonresonant dielectric defect with finite permittivity contrast described by an effective polarizability. In both cases, physically admissible reciprocal perturbations can be constructed for which the leading backward-scattering amplitude is nonzero. Reciprocity alone therefore does not guarantee reflectionless propagation against arbitrary reciprocal disorder.

This no-go result does not preclude strong suppression of backscattering for suitably restricted classes of perturbations. In Sec.~III, we show that the leading-order backscattering expressions derived for these two disorder limits can also provide constructive guidance for suppressing reflection by engineering the local polarization structure of the guided mode. We develop this approach through two examples: evanescent-field polarization engineering and a polarization-selective gap induced by structural circular dichroism. In Sec.~IV, we revisit representative reciprocal valley-Hall and spin-Hall photonic topological waveguides, examine the mechanisms that allow coupling between their counter-propagating modes, and use full-wave simulations to demonstrate substantial backscattering induced by local defects and sharp bends. Together, these results distinguish universal backscattering immunity from disorder-specific suppression and clarify the scope of robust transport achievable in reciprocal photonic waveguides.

\section{General argument}
Here, we present a simplified version of the general proof to highlight the underlying physical intuition of why the universal reflectionless transport is impossible in the presence of reciprocity. Let us consider a linear reciprocal waveguide such as vacuum or fiber that is continuous translational invariant along the propagation direction, or a photonic crystal waveguide with discrete translational symmetry. The materials made of the waveguide can be macroscopically described by the full constitutive matrix $\mathcal{C}=\left(\begin{array}{ll} \boldsymbol{\varepsilon} & \boldsymbol{\xi} \\ \boldsymbol{\zeta} & \boldsymbol{\mu} \end{array}\right)$ where $\boldsymbol{\varepsilon}$ and $\boldsymbol{\mu}$ denote the permittivity and permeability tensors, respectively, while $\boldsymbol{\xi}$ and $\boldsymbol{\zeta}$ describe the magnetoelectric coupling~\cite{kongElectromagneticWaveTheory2000}. In this settings, the eigenmodes of the system can be labelled by a bloch wavevector $\mathbf{k}$ along $x$ direction according to the Bloch theorem \cite{ashcroftSolidStatePhysics2012,joannopoulosPhotonicCrystalsMolding2011}. Let 
$\boldsymbol{\Psi}_{\mathbf{k}}
=
\begin{pmatrix}
\mathbf E_{\mathbf{k}}\\
\mathbf H_{\mathbf{k}}
\end{pmatrix}$ denote a forward-moving Bloch mode at angular frequency $\omega$ associated with a possibly complex Bloch wavevector $\mathbf{k}$. Then the reciprocity ensures the existence of its backward-moving counterpart $\mathbf{\Psi}_{-\mathbf{k}}$ at the same frequency, associated with the opposite Bloch wavevector \(-\mathbf{k}\) \cite{SupplementaryMaterial}.

For simplicity, the main-text proof is formulated for the waveguiding systems with relative permeability $\boldsymbol{\mu}_r=\mathbf{I}$ and vanishing electric--magnetic coupling $\boldsymbol{\xi}=\boldsymbol{\zeta}=\mathbf{0}$, while allowing a general spatially varying permittivity tensor $\boldsymbol{\varepsilon}(\mathbf r)$. We consider the localized disorder described by a position-dependent permittivity perturbation $\Delta\boldsymbol{\varepsilon}(\mathbf{r})$, such that the scattering analysis can be formulated solely in terms of the modal electric fields $\mathbf{E}_{\pm\mathbf{k}}$. These settings apply to many of the most common experimentally relevant optical and electromagnetic waveguides platform. The extension to general constitutive media $\mathcal{C}(\mathbf{r})$ and perturbations of the full constitutive matrix $\Delta\mathcal{C}(\mathbf{r})$ is given in the Supplementary Material. Considering a generic compactly supported dielectric perturbation localized around $\mathbf r$, we focus on two types that are common in practice: A Type-1 scatterer has a finite spatial extent, with $|\Delta\varepsilon_{ij}(\mathbf r)|$ spatially uniform and sufficiently small throughout the perturbed region, whereas a Type-2 scatterer has a small spatial extent, with a spatially uniform permittivity contrast of an admissible finite magnitude throughout the perturbed region. Typical examples of the Type-1 include weak material variations arising from fabrication, composition fluctuations, and thermorefractive-index variations, whereas the Type-2 includes nanoscale dielectric inclusions and subwavelength sidewall-roughness disorder. Suppose that the forward guided Bloch mode $\mathbf E_{\mathbf k}$ is incident on such a disorder, the leading-order backscattering amplitude into its reciprocal backward counterpart $\mathbf E_{-\mathbf k}$ is proportional to the following overlap integral \cite{SupplementaryMaterial}:
\begin{equation}
    c_{-\mathbf{k} \rightarrow \mathbf{k}} 
    \propto 
    \left \langle{
    \mathbf E_{\mathbf k}^*}
    \right|
    \hat{\Delta}
    \left|
    \mathbf E_{\mathbf k}
    \right\rangle \label{eq:2.1}
\end{equation}
where the $\hat{\Delta}$ is a linear operator representing the perturbation. If the waveguide system is Hermitian and we choose $\mathbf{E}_{\mathbf{-k}}=\mathbf E_{\mathbf k}^*$, then the right hand side (RHS) of Eq.~(\ref{eq:2.1}) can be written as $\langle \mathbf E_{-\mathbf k}| \hat{\Delta}|\mathbf E_{\mathbf k}\rangle$. The action of the perturbation on the incident mode, $\hat{\Delta}|\mathbf E_{\mathbf k}\rangle$, can be represented as an effective polarization-current source $\mathbf J$ that is linear in $\mathbf E_{\mathbf k}$. The RHS of Eq.~(\ref{eq:2.1}) therefore represents the overlap of the reciprocal backward mode with this effective source and hence the excitation amplitude of the backward-propagating mode $\mathbf E_{-\mathbf k}$. The exact expression for this amplitude follows from the reciprocity theorem~\cite{SupplementaryMaterial,collinFieldTheoryGuided1991,shun-lienchuangCoupledModeFormulation1987,lecampTheoreticalComputationalConcepts2007a}. For the backscattering induced by a Type-1 scatterer, $\hat{\Delta}$ is linearly proportional to the permittivity perturbation, $\hat{\Delta}\sim\Delta\boldsymbol{\varepsilon}$, within the first Born approximation, where the small-magnitude condition requires $|\Delta\varepsilon_{ij}(\mathbf r)|$ to be sufficiently small that multiple scattering within the perturbed region is negligible~\cite{jackson1998classical,sakuraiModernQuantumMechanics2020,chewWavesFieldsInhomogeneous1995,habashyBornRytovApproximations1993}. The corresponding effective source is a distributed polarization current $\mathbf J=-\mathrm{i}\omega\Delta\boldsymbol{\varepsilon}\mathbf E_{\mathbf k}$ over the perturbed region~\cite{jackson1998classical,chewWavesFieldsInhomogeneous1995,snyderOpticalWaveguideTheory1984,barwiczThreedimensionalAnalysisScattering2005}. For a Type-2 scatterer, the leading-order scattering response can be represented by an effective electric-dipole polarizability tensor, $\hat{\Delta}\sim\boldsymbol{\alpha}_{\mathrm{eff}}$, within the quasistatic electric-dipole approximation, where the small-spatial-extent condition requires the characteristic size of the scatterer to be much smaller than the characteristic length scales over which the exciting and internal fields vary, such that the electric-dipole response dominates over higher-order multipolar contributions~\cite{jackson1998classical,bohren1998absorption,novotny2012principles}. The corresponding effective source is point electric-dipole current $\mathbf J=-\mathrm{i}\omega\boldsymbol{\alpha}_{\mathrm{eff}}\mathbf E_{\mathbf k}$~\cite{jackson1998classical,johnson2005roughness,kitaAreSlotSubwavelength2018}.

We show that a truly backscattering-immune waveguide is fundamentally impossible by establishing that the coupling coefficient \(c_{-\mathbf{k}\rightarrow\mathbf{k}}\) in Eq.~(\ref{eq:2.1}) cannot vanish identically for arbitrary perturbations in a linear reciprocal waveguide. For the guided mode in the waveguide, there necessarily exists a finite region of the waveguide over which \(\mathbf E_{\mathbf k}(\mathbf r)\) is not identically zero. A localized disorder characterized by \(\hat{\Delta}\) can therefore be introduced within such a region, and complete immunity against arbitrary disorder would require the overlap in Eq.~(\ref{eq:2.1}) to vanish for every physically admissible realization of \(\hat{\Delta}\). For the reciprocal perturbations considered here, reciprocity requires the permittivity perturbation \(\Delta\boldsymbol{\varepsilon}(\mathbf r)\) to be transpose symmetric, and the corresponding linear electric-response operator \(\hat{\Delta}\) inherits the same transpose symmetry \cite{SupplementaryMaterial}. Reciprocity imposes no further restriction on its tensorial structure. Consequently, the variations in the tensor components of the \(\Delta\boldsymbol{\varepsilon}(\mathbf r)\), in the disorder geometry and position could generate a continuous family of transpose-symmetric, potentially anisotropic electric-response operators $\hat{\Delta}$. We would demonstrate below, with the Type-1 and Type-2 disorder, how the freedom in the tensorial structure of $\hat{\Delta}$ allowed by the reciprocity eventually precludes the universal reflectionlessness.

For Type-1 disorder, the tensorial freedom of $\hat{\Delta}$ is inherited directly from the local permittivity perturbation $\Delta\boldsymbol{\varepsilon}(\mathbf r)$. Under the first-Born approximation, the integrand of the overlap integral in Eq.~(\ref{eq:2.1}) is:
\begin{equation}
    \mathbf{E}_{\mathrm{k}}(\mathbf{r}) \cdot \Delta \varepsilon(\mathbf{r}) \mathbf{E}_{\mathrm{k}}(\mathbf{r}) \label{eq:2.2}
\end{equation}
Weak material fluctuations --- for example, local composition or dopant fluctuations, thermorefractive fluctuations, and strain-induced photoelastic variations ---  can in general perturb different components of the permittivity tensor by different amounts, while preserving $\Delta \boldsymbol{\varepsilon}^{\mathrm T}=\Delta \varepsilon$. Around a generic position \(\mathbf r\) where \(\mathbf E_{\mathbf k}(\mathbf r)\neq\mathbf 0\), one can always identify a weak, compactly supported, transpose-symmetric perturbations for which Eq.~(\ref{eq:2.2}) is locally nonzero at $\mathbf{r}$ and the resulting spatial overlap in Eq.~(\ref{eq:2.1}) does not vanish, yielding a nonzero back-scattering amplitude. Explicit constructions for perturbations located within the constituent material regions and in the vicinity of their interfaces that produces nonvanishing back-scattering are given in the Supplementary Material~\cite{SupplementaryMaterial}.

Type-2 disorder provides a complementary route to the same tensorial freedom through geometric anisotropy alone. For waveguides composed of isotropic constituent materials, an experimentally common Type-2 disorder is described by scalar permittivity variations $\Delta\varepsilon$. We consider two representatives geometries of them, that are commonly used to model localized bulk and interfacial imperfections. Within the bulk of the constituent material, we consider a small spheroidal disorder, corresponding experimentally to localized voids, precipitates, or embedded particles. At an interface between two constituent materials, we consider a localized protrusion or indentation with a small semi-spheroidal profile whose symmetry axis is normal to the local interface \footnote{\label{Type2InterfacialGeometry} For the interfacial disorder, the characteristic disorder size is assumed to be much smaller than the spatial-variation scales of the tangential electric field and the normal electric displacement of the incident mode along the interface, so that the local modal excitation can be treated as approximately uniform over the defect. The disorder size is also taken to be much smaller than the local radius of curvature of the unperturbed interface, such that the background interface is locally planar on the disorder scale. The locally smoothness ($C^2$ smooth) assumption on the interfacial profile excludes the nontrivial essential-spectrum regimes associated with corners and edges~\cite{BonnetBenDhia2012,PerfektPutinar2017}.}, as may arise from fabrication-induced surface or sidewall roughness \cite{kitaAreSlotSubwavelength2018}. For an appropriate finite contrast in permittivity $\Delta\varepsilon$ \footnote{\label{Type2Regularity}
For the two representative Type-2 disorder considered here, the finite material contrast $\Delta\varepsilon$ is restricted to values away from the exceptional plasmonic contrasts at which the quasistatic response becomes singular or resonant. In particular, we exclude the electrostatic surface-plasmon condition $\Delta\varepsilon=-2\varepsilon_b$ where the $\varepsilon_b$ is the unperturbed background permittivity in the disorder region, for both the bulk and interfacial disorder. In addition, the discrete plasmonic contrasts associated with localized quasistatic resonances of the specific disorder geometry are likewise excluded~\cite{AmmariKang2007,BonnetBenDhia2012,AndoKang2016,Mayergoyz2005}. Complex permittivities, including loss and gain, are otherwise allowed.
}, the scattering response of the two disorders discussed above would be dominated by the electric-dipole contribution. The backscattering Eq.~(\ref{eq:2.1}) induced by a disorder at $\mathbf{r}$ would thereby reduces to:
\begin{equation}
    \mathbf{E}_{\mathbf{k}}(\mathbf{r})\cdot
    \boldsymbol{\alpha}_{\mathrm{eff}}(\mathbf{r})
    \mathbf{E}_{\mathbf{k}}(\mathbf{r}),
    \label{eq:2.3}
\end{equation}
Varying the available geometric degrees of freedom of the bulk and interfacial disorders prescribed above \footnote{\label{Type2GeometryVariation}
Here, the aspect ratio denotes the ratio of the symmetry-axis to transverse semiaxes of the spheroidal profile, while the orientation specifies the direction of its symmetry axis relative to the local modal-field polarization. For the interfacial semi-spheroid, the orientation is fixed by the local interface normal and only the aspect ratio is varied; for the embedded spheroid, both quantities may be varied. For plasmonic material contrasts, discrete aspect ratios at which localized quasistatic plasmon resonances occur are excluded \cite{Berreman1970,Moskovits1985,Mayergoyz2005}. 
}, through changes in their aspect ratio or orientation relative to the polarization direction of $\mathbf{E}_{\mathbf{k}}(\mathbf{r})$, generates continuous families of transpose-symmetric, generally anisotropic effective electric-dipole polarizability tensors $\boldsymbol{\alpha}{\mathrm{eff}}$, which can correspond to distinct, noncollinear electric-dipole response. Consequently, at a generic position $\mathbf r$ where $\mathbf E_{\mathbf k}(\mathbf r)\neq\mathbf 0$, this geometric freedom ensures the existence of an admissible Type-2 disorder for which Eq.~(\ref{eq:2.3}) is nonzero, thereby yielding nonvanishing backscattering, as demonstrated in the Supplementary Material.

We make several remarks on scope of the above proof. The proof is formulated at a fixed operating frequency and applies to linear, time-invariant reciprocal waveguides subject to static disorder. Within this setting, the macroscopic constitutive responses of both the unperturbed waveguide and the disorder can complex, anisotropic, dispersive, and non-Hermitian. Spatially nonlocal response, including spatial dispersion that may become relevant in effective-medium descriptions of metamaterials~\cite{AgranovichGinzburg1984,Silveirinha2007}, are not considered in the present analysis. The proof of the absence of universal backscattering immunity is based on a sufficient condition: the coupling between a pair of counter-propagating guided modes cannot vanish for every admissible reciprocal disorder realization. We restrict attention to operating points at which these guided channels are well defined and possess finite modal fields and fluxes. Throughout the proof, $\mathbf E_{\mathbf k}$ and $\mathbf E_{-\mathbf k}$ denote a pair of distinct counter-propagating guided modes carrying nonzero longitudinal modal flux. Radiation modes and modal thresholds, including cutoffs and band edges at which such propagating channels cease to be distinct or carry vanishing flux, are not considered.

The proof with the Type-1 perturbation is built upon the assumption that the full-Maxwell scattering problem is well posed at the operating frequency and remains regular under sufficiently weak localized perturbations, such that the outgoing-wave condition determines a unique scattered field with finite outgoing modal amplitudes. Exceptional cases in which an arbitrarily weak perturbation drives the system to a singular scattering configuration, for example by reaching a lasing threshold, fall outside this construction and are discussed in the \cite{SupplementaryMaterial}. Under this regularity condition, a sufficiently weak, compactly supported reciprocal perturbation can be treated within the first Born approximation. For the effective-polarizability representation of a Type-2 perturbation, the associated quasistatic boundary-value problem is required to remain regular~\cite{AmmariKang2007,BonnetBenDhia2012,bohren1983absorption,Mayergoyz2005}. For a non-epsilon-near-zero (non-ENZ) scalar background, this condition can be satisfied by an appropriate choice of the finite permittivity contrast and disorder geometry, away from the quasistatic singular or resonant configurations as detailed above. The presence of exact-ENZ regions elsewhere in the waveguide does not obstruct the construction, since the disorder may instead be placed within a regular non-ENZ constituent region. As shown in the Supplementary Material~\cite{SupplementaryMaterial}, any power-carrying guided mode with distinct counterpropagating partners in a local piecewise-scalar dielectric waveguide must have nonzero modal support in at least one such non-ENZ region. The Type-2 construction can therefore always be carried out within a regular local background even when exact-ENZ constituents are present elsewhere in the structure.

The fundamental origin of this limitation can be understood by considering the conditions required for disorder-robust reflectionless transport in physical systems supporting counter-propagating partners, including time-reversal-invariant electronic systems and reciprocal photonic systems. Here, we consider disorder-independent mechanisms for maintaining reflectionless transport in the presence of localized perturbations of finite spatial extent. By disorder-independent, we mean that backscattering remains forbidden throughout a prescribed physical class of disorder, without requiring individual disorder realizations to be specifically engineered to cancel reflection. Strict backscattering immunity within such a disorder class requires an exact constraint that enforces vanishing forward-to-backward scattering in the full scattering problem for every allowed perturbation. Symmetry provides a particularly important mechanism for enforcing such a constraint, by imposing a selection rule that forbids coupling between the counter-propagating channels.

For electronic system, the intrinsic spin-$\tfrac12$ degree of freedom (DoF) provides a natural realization of such a selection rule. When the two counter-propagating channels carry different intrinsic-spin quantum numbers, any spin-independent disorder cannot couple the two channels. The corresponding protected class already encompasses a broad class of spatial and orbital disorders. For example, a scalar disorder potential $V(\mathbf{r})=v(\mathbf{r}) I_{\text {spin}}$ can vary arbitrarily while acting trivially on the spin DoF and therefore would not induce backscattering between the two counter-propagating channels. Importantly, the spin degree of freedom is intrinsic to the electron rather than encoded in the spatial or material structure of the underlying system. Preserving the associated spin selection rule therefore does not require the disorder to maintain any specially engineered spatial or material property of the host system. More generally, fermionic time-reversal symmetry $\Theta_{\mathrm e}$, with $\Theta_{\mathrm e}^{2}=-1$, provides a stronger selection rule: two counter-propagating states related by $\Theta_{\mathrm e}$ form a Kramers doublet, and elastic backscattering between them is forbidden by any static time-reversal-preserving perturbations $\Theta_{\mathrm e}V\Theta_{\mathrm e}^{-1}=V$. Kramers protection thus permits backscattering immunity throughout the full class of static perturbations allowed by the fundamental time-reversal constraint of the electronic system.

For photonic systems, the situation is fundamentally different. Although the photons possess intrinsic spin-$1$, the physical spin states are helicity states defined with respect to the propagating momentum. Consequently, in a generic structured photonic system, the intrinsic photon spin does not furnish a conserved quantum number. The two counter-propagating channels therefore cannot be assigned opposite intrinsic-spin quantum numbers in the same manner as in a structured electronic system with conserved intrinsic electron spin. In addition, the bosonic time reversal satisfies $\Theta_{\mathrm{ph}}^2=+1$ so two counter-propagating states related by \(\Theta_{\mathrm{ph}}\) do not form a Kramers doublet. Therefore, unlike in the electronic case, photonic time-reversal symmetry does not provide backscattering immunity throughout the full class of time-reversal-preserving perturbations in the photonic system. 

Since neither intrinsic photon spin nor intrinsic time-reversal symmetry provides the required selection rule in the electromagnetic system, any exact constraint enforcing backscattering immunity must instead be engineered through the hosting electromagnetic media. Within the local macroscopic Maxwell description considered here, such engineering is implemented through the spatial and tensorial constitutive parameters $\mathcal{C}(\mathbf{r}, \omega)=\left(\begin{array}{ll}
\boldsymbol{\varepsilon} & \boldsymbol{\xi} \\
\boldsymbol{\zeta} & \boldsymbol{\mu}
\end{array}\right)$. This immediately introduces a vulnerability that is not inherent to the selection rules built upon intrinsic electron spin and fermionic Kramers protection in the electronic systems discussed above: the same constitutive parameters through which the constraint is engineered can also be directly modified by fabrication imperfections and environmental perturbations. Fabrication imperfections, material-composition fluctuations, and other common form of disorders can all generate perturbations $\mathcal{C}(\mathbf{r}, \omega) \rightarrow \mathcal{C}(\mathbf{r}, \omega)+\Delta \mathcal{C}(\mathbf{r}, \omega)$. These disorders remain fully compatible with the Lorentz reciprocity since ordinary fabrication and material disorder does not introduce magneto-optic effects or time modulation. Reciprocity requires the perturbation \(\Delta\mathcal C\) to preserve the reciprocal symmetry relations among the constitutive tensors, but does not require the perturbation to preserve the additional structure engineered in $\mathcal C$ to enforce backscattering suppression. The corresponding backscattering-suppression constraint therefore generally defines a more restricted subset of the reciprocal constitutive-parameter space, and reciprocity alone does not require a perturbation to remain within this subset. A reciprocal perturbation can thus preserve reciprocity while violating the additional constraint and thereby inducing coupling between the counter-propagating channels.

A prominent realization of such engineered constraints is provided by reciprocal photonic topological insulators inspired by the electronic quantum spin Hall (QSH) effect, in which an artificial pseudospin degree of freedom is constructed to emulate the two spin channels of the electronic system. The resulting suppression of backscattering, however, relies on additional structure beyond reciprocity, and physically ordinary reciprocal perturbations can violate the corresponding pseudospin constraint. In the Wu--Hu photonic-crystal model, the pseudospin states are constructed from the \(p_{\pm}\)- and \(d_{\pm}\)-like modes associated with the designed \(C_{6}\) crystalline symmetry; a local reciprocal geometrical perturbation that breaks the relevant crystalline structure can therefore mix the pseudospin sectors \cite{wu2015scheme}. In coupled-resonator optical waveguide (CROW) realizations, clockwise and counter-clockwise resonator modes play the role of the two pseudospins, and protection requires these modes to remain decoupled; reciprocal backscattering produced, for example, by resonator sidewall disorder directly couples them \cite{hafezi2011robust,hafezi2013imaging}. In bianisotropic TE/TM constructions, the pseudospins are formed from hybridized TE and TM modes whose degeneracy and decoupling rely on an engineered balance between the electric and magnetic responses, commonly expressed through an electromagnetic-duality condition; reciprocal perturbations that disturb this balance can likewise mix the two pseudospins \cite{khanikaev2013photonic,ma2015guiding}. These examples illustrate that the symmetry-enforced selection rules of reciprocal QSH-inspired photonic systems are protected only within the subset of reciprocal perturbations that preserves the additional structure on which the pseudospin construction relies.

\section{Constructive routes to backscattering suppression}

We have demonstrated that disorder-independent backscattering immunity is fundamentally unattainable in the reciprocal optical system because the tensorial product Eq.~(\ref{eq:2.2}) and Eq.~(\ref{eq:2.3}) cannot be identically zero as long as $\mathbf{E}_{\mathbf{k}}$ is non-trivial. The same identities, however, also provide a constructive principle in the reciprocal waveguide for suppressing the backscattering caused by a broad class of experimentally-relevant scatterers. Two design routes follow immediately. The first is to reduce the amplitude of the unperturbed Bloch mode in regions where disorder is expected to occur, such as waveguide boundaries \cite{melatiRealPhotonicWaveguides2014}. In this case, any localized Type-1 or Type-2 perturbation interacts only weakly with the mode, and the resulting reflection is correspondingly suppressed. This is the conventional strategy underlying many approaches to reducing scattering loss from fabrication-induced surface roughness in dielectric waveguides \cite{melatiRealPhotonicWaveguides2014,huLowlossHighindexcontrastPlanar2007,bockSubwavelengthGratingPeriodic2010,johnsonLowlossAsymptoticallySinglemode2001,yehTheoryBraggFiber1978}

The second route is built upon the vectorial nature of the electromagnetic field. Instead of reducing the modal intensity, one engineers the local polarization of the Bloch mode $\mathbf{E}_{\mathbf{k}}$ such that the tensorial product Eqs.~(\ref{eq:2.2}) and~(\ref{eq:2.3}) vanish over the perturbed region. The leading-order backscattering amplitude can then be suppressed at finite field intensity. Specifically, Eq.~(\ref{eq:2.3}) represents the leading electric-dipole contribution in the electrically small-scatterer limit, whereas Eq.~(\ref{eq:2.2}) corresponds to the first-Born contribution in the weak-contrast limit. The cancellation Eq.~(\ref{eq:2.2}) and Eq.~(\ref{eq:2.3}) of does not imply strictly zero reflection for a real-world scatterer of finite size and finite material contrast, since multipole scattering or multiple scattering processes might still retains. Nevertheless, for many practical disorders whose response is dominated by the electric-dipolar or single-scattering contribution, suppressing this leading term make the reflection approach zero in the corresponding asymptotic limits. Such suppression is therefore best understood as a disorder-specific polarization-selection effect enabled by the vectorial nature of the electromagnetic field, rather than as universal topological or symmetry protection.

To make this route explicit, we specialize to a lossless, isotropic dielectric waveguide in a frequency range supporting a single pair of reciprocal guided modes. The local electric field of the forward Bloch mode is decomposed in an orthonormal basis $\{\hat{\mathbf e}_1,\hat{\mathbf e}_2,\hat{\mathbf e}_3\}$ as
\begin{equation}
    \mathbf{E}_{\mathbf{k}}(\mathbf r)
    =
    E_1(\mathbf r)\hat{\mathbf e}_1
    +
    E_2(\mathbf r)\hat{\mathbf e}_2
    +
    E_3(\mathbf r)\hat{\mathbf e}_3 ,
    \label{eq:3.1}
\end{equation}
where $E_1$, $E_2$, and $E_3$ are generally complex. In the same basis, the components of the effective polarizability tensor are defined by $\alpha_{ij}^{\mathrm{eff}}, \{i,j\}\in\{1,2,3\}$. We first consider a Type-2 scatterer centered at $\mathbf r_0$. If two basis directions can be chosen such that the corresponding cross-polarizability terms vanish, $\alpha_{12}^{\mathrm{eff}}=\alpha_{21}^{\mathrm{eff}}=0$, Eq.~(\ref{eq:2.3}) can be written as
\begin{equation}
    \mathbf E_{\mathbf k}\cdot
    \boldsymbol{\alpha}_{\mathrm{eff}}
    \mathbf E_{\mathbf k}
    =
    \alpha_{11}^{\mathrm{eff}}E_1^2
    +
    \alpha_{22}^{\mathrm{eff}}E_2^2
    +
    \mathcal R(E_3),
    \label{eq:3.4}
\end{equation}
where the explicit dependence on $\mathbf r_0$ has been omitted for notational simplicity, and $\mathcal R(E_3)$ denotes the remaining terms that are linear or quadratic in $E_3$. For a lossless dielectric scatterer with real, same-sign $\alpha_{11}^{\mathrm{eff}}$ and $\alpha_{22}^{\mathrm{eff}}$, the first two terms in Eq.~(\ref{eq:3.4}) cancel when
\begin{equation}
    \frac{E_1}{E_2}
    =
    \pm i
    \sqrt{
    \frac{\alpha_{22}^{\mathrm{eff}}}
         {\alpha_{11}^{\mathrm{eff}}}
    } .
    \label{eq:3.5}
\end{equation}
The two field components are therefore required to be in quadrature, with a relative phase of $\pm\pi/2$, while their amplitude ratio is determined by the corresponding effective polarizabilities. We next consider a Type-1 scatterer. For a weak permittivity perturbation in an isotropic dielectric medium, the constitutive perturbation takes the scalar form $\Delta\boldsymbol{\varepsilon}(\mathbf r)\rightarrow \Delta\varepsilon(\mathbf r)$. Equation~(\ref{eq:2.2}) therefore reduces locally to
\begin{equation}
    \mathbf E_{\mathbf k}(\mathbf r)\cdot
    \Delta\boldsymbol{\varepsilon}(\mathbf r)
    \mathbf E_{\mathbf k}(\mathbf r)
    =
    \Delta\varepsilon(\mathbf r)
    \left[
    E_1^2(\mathbf r)
    +
    E_2^2(\mathbf r)
    \right]
    +
    \mathcal R(E_3)
    \label{eq:3.6}
\end{equation}
The contribution from the first term vanishes when 
\begin{equation}
    \frac{E_1}{E_2}
    =
    \pm i
    \label{eq:3.7}
\end{equation}
corresponding to equal amplitudes and a relative phase of $\pm\pi/2$, which essentially requires the $\mathbf{E}_{\mathbf k}(\mathbf r)$ to be circularly polarized over the perturbed region. 

The relations in Eqs.~(\ref{eq:3.5}) and~(\ref{eq:3.7}) both require the electric field of the forward guided mode at the position of the local disorder to be elliptically polarized in the $\{\hat{\mathbf e}_1,\hat{\mathbf e}_2\}$ plane. This establishes a direct connection between local polarization handedness and the suppression of the leading-order backscattering, suggesting that we can suppress the reflection induced by the given scatterer by engineering the polarization structures of the guided modes. Realizing the cancellation conditions in Eqs.~(\ref{eq:3.5}) and~(\ref{eq:3.7}) for the full three-dimensional overlap additionally requires the $E_3$-dependent contribution to vanish or remain non-dominant relative to the contributions associated with $E_1$ and $E_2$, which is not strict requirement in many waveguide platform as discussed in \cite{SupplementaryMaterial}.

We now give two concrete examples of this polarization engineering mechanism for the suppression of the reflection induced by the scatterer in a reciprocal waveguide. 

\subsection{Evanescent wave engineering}
The first example exploits the intrinsic polarization structure of an evanescent field. For an evanescent wave in a lossless isotropic medium propagating along $\hat{\mathbf e}_1$ and decaying along $\hat{\mathbf e}_2$, the complex wave vector can be written as $\mathbf{k}_{\mathrm{ev}}=k\hat{\mathbf e}_1+\mathrm{i}\kappa\hat{\mathbf e}_2$. The transversality condition $\mathbf{k}_{\mathrm{ev}}\cdot\mathbf E=0$ gives $E_1/E_2=\pm\mathrm{i}\kappa/k$, so the two field components are intrinsically in quadrature and the local field is elliptically polarized. This polarization handedness underlies the well-known spin--momentum locking of evanescent waves \cite{rodriguez-fortunoNearFieldInterferenceUnidirectional2013,vanmechelenUniversalSpinmomentumLocking2016a,bliokhSpinOrbitInteractions2015,oconnorSpinOrbitCoupling2014,petersenChiralNanophotonicWaveguide2014}. Such evanescent-field polarization occurs broadly across many guided-wave platforms, including dielectric waveguides, fibers, and surface-wave waveguides \cite{vanmechelenUniversalSpinmomentumLocking2016a}. A particularly useful property of such an evanescent field is that its local polarization structure is fixed by the transversality condition. Importantly, within a homogeneous evanescent-field region, the ratio $E_1/E_2$ is position independent. Hence, once Eq.~(\ref{eq:3.5}) is satisfied for a prescribed Type-2 scatterer, the leading electric-dipole back-scattering remains suppressed under translations of the scatterer within that region, as long as its relevant effective polarizability $\boldsymbol{\alpha}_{\mathrm{eff}}$ is unchanged.

\begin{figure}[h]
    \centering
    \includegraphics[width=\linewidth,height=0.6\textheight,keepaspectratio]{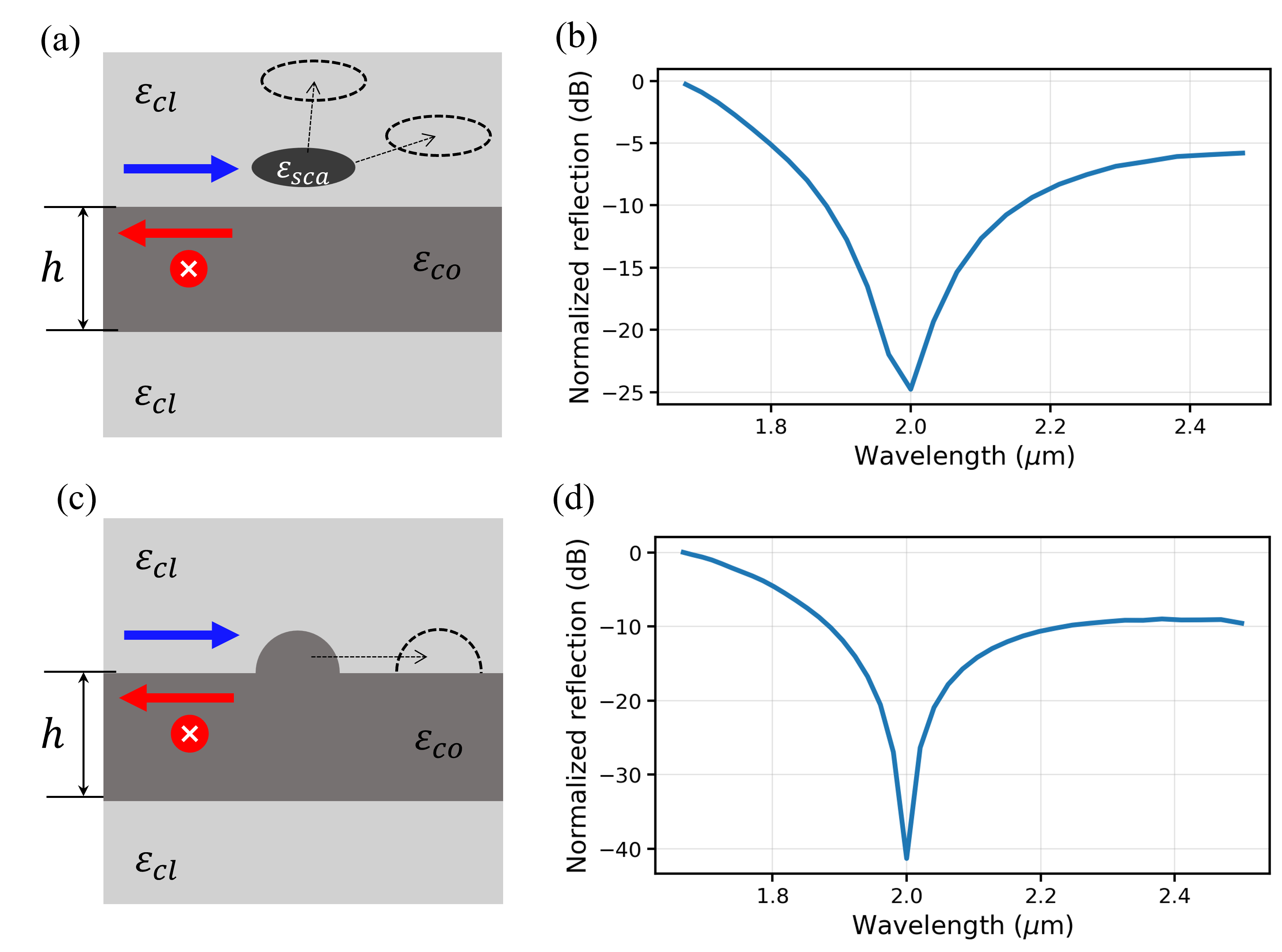}
    \caption{Simulation schematics and reflection spectra. (a) Schematic of a small elliptical dielectric scatterer embedded in the cladding of the slab waveguide. (b) Normalized reflection spectrum for the elliptical scatterer in (a). (c) Schematic of a mirror-symmetric raised bump at the core--cladding interface of the slab waveguide. (d) Normalized reflection spectrum for the interfacial scatterer in (c).}
    \label{fig:2.1.1}
\end{figure}

The simplest 2D slab waveguide provides a particularly transparent platform for engineering the evanescent field to realize the cancellation condition in Eq.~(\ref{eq:3.5}). We consider the fundamental TM$_0$ mode of a symmetric dielectric slab with a core of thickness $h$ centered at $z=0$ and cladding occupying $|z|>h/2$. The core and cladding permittivities are denoted by $\varepsilon_{\mathrm{co}}$ and $\varepsilon_{\mathrm{cl}}$, respectively, with $\rho=\varepsilon_{\mathrm{co}}/\varepsilon_{\mathrm{cl}}$. Taking $x$ as the propagation direction and $z$ as the transverse confinement direction, the evanescent field in the cladding satisfies $E_x/E_z=-\mathrm{i}\kappa/k$, where $k$ is the propagation constant and $\kappa=(k^2-\varepsilon_{\mathrm{cl}}k_0^2)^{1/2}>0$ is the transverse decay constant. The polarization ratio $\eta\equiv|E_x/E_z|=\kappa/k$ can therefore be tuned through the slab-waveguide dispersion. We consider two representative Type-2 scatterers, as shown in Fig.~\ref{fig:2.1.1}(a,c): a small elliptical scatterer embedded in the cladding and a mirror-symmetric raised bump at the core--cladding interface. The former represents a particulate dielectric particle or inclusion in the cladding, while the latter could be a common fabrication imperfection induced localized protrusion at the core--cladding boundary. Their spatial extents are small compared with the characteristic variation scales of the unperturbed modal field, so that their leading responses are described by effective electric-dipole polarizability tensors. For the interfacial bump, the corresponding small-size condition in the normal direction is defined with respect to the variation of $D_z$, rather than the discontinuous $E_z$ \cite{johnson2005roughness}. Owing to the mirror symmetries of the corresponding quasistatic boundary-value problems, $\boldsymbol{\alpha}_{\mathrm{eff}}$ is diagonal in the $\{\hat{\mathbf x},\hat{\mathbf z}\}$ basis for both scatterers; the relevant components are obtained analytically for the elliptical scatterer and numerically for the surface bump \cite{jackson1998classical,bohren1998absorption,johnson2005roughness,khalatpourRoughnessLimitedPerformanceUltraLowLoss2026}.

For either Type-2 scatterer, Eq.~(\ref{eq:3.5}) reduces to
\begin{equation*}
    \eta^2
    =
    \left|\frac{E_x}{E_z}\right|^2
    =
    \frac{\alpha_{zz}^{\mathrm{eff}}}
         {\alpha_{xx}^{\mathrm{eff}}}.
\end{equation*}
The required value of $\eta$ is therefore fixed by the ratio of the two diagonal effective-polarizability components. Since $\eta=|E_x/E_z|=\kappa/k$ for the evanescent field in the cladding, the slab geometry required to realize this polarization ratio follows directly from the modal dispersion. The normalized propagation constant is
\begin{equation*}
    b=
    \frac{k^2-\varepsilon_{\mathrm{cl}}k_0^2}
         {(\varepsilon_{\mathrm{co}}-\varepsilon_{\mathrm{cl}})k_0^2}=
    \frac{\eta^2}
         {(\rho-1)(1-\eta^2)}.
\end{equation*}
For the even TM$_0$ mode of the symmetric slab, the dispersion relation gives
\begin{equation}
    V=
    \frac{1}{\sqrt{1-b}}
    \tan^{-1}\!\left[
        \rho\sqrt{\frac{b}{1-b}}
    \right],
    \quad
    V=
    \frac{k_0h}{2}
    \sqrt{\varepsilon_{\mathrm{co}}-\varepsilon_{\mathrm{cl}}},
    \label{eq:3.8}
\end{equation}
where $k_0=2\pi/\lambda_0$. Thus, for a prescribed Type-2 scatterer, Eq.~(\ref{eq:3.5}) fixes the required polarization ratio $\eta$, while Eq.~(\ref{eq:3.8}) determines the corresponding slab thickness at the target wavelength.
For numerical verification, we set $\varepsilon_{\mathrm{co}}=8$ and $\varepsilon_{\mathrm{cl}}=1$ thus $\rho=\varepsilon_{\mathrm{co}}/\varepsilon_{\mathrm{cl}}=8$ and determine the slab thickness for each scatterer from Eqs.~(\ref{eq:3.5}) and~(\ref{eq:3.8}), such that the cancellation condition is satisfied at the target wavelength $2~\mu\mathrm{m}$. For the elliptical scatterer, we take an aspect ratio equal to $4$, with the major and minor axes along $x$ and $z$, respectively, and $\varepsilon_{\mathrm{sca}}/\varepsilon_{\mathrm{cl}}=16$. The quasistatic polarizability gives $\alpha_{zz}^{\mathrm{eff}}/\alpha_{xx}^{\mathrm{eff}}=1/3.25$ \cite{bohren1998absorption}, corresponding to $\eta=0.555$. Equation~(\ref{eq:3.8}) then gives $V=1.161$ and $h=0.28~\mu\mathrm{m}$ for the structure in Fig.~\ref{fig:2.1.1}(a). For the mirror-symmetric raised bump, the effective polarizabilities are $\alpha_{xx}^{\mathrm{eff}}=[(\varepsilon_{\mathrm{co}}+\varepsilon_{\mathrm{cl}})/2]\alpha(\rho)$ and $\alpha_{zz}^{\mathrm{eff}}=\varepsilon_{\mathrm{cl}}^2\gamma(\rho)$ within the effective-current formulation \cite{johnson2005roughness}. For the semi-hemisphere bump geometry considered here, $\alpha(8)=0.829$ and $\gamma(8)=1.486$, yielding $\alpha_{zz}^{\mathrm{eff}}/\alpha_{xx}^{\mathrm{eff}}=0.398$ and hence $\eta=0.631$. The corresponding design parameters are $V=1.263$ and $h=0.3~\mu\mathrm{m}$ for the structure in Fig.~\ref{fig:2.1.1}(c). Both designs satisfy $V<\pi/2$ and therefore remain below the cutoff of the first higher-order mode in the TM polarization sector.

Figures~\ref{fig:2.1.1}(b,d) show the reflected power into the reciprocal TM$_0$ mode for the two optimized structures. Both spectra exhibit a pronounced reflection minimum near the polarization-matching condition, where the leading-order backward coupling is suppressed. For the elliptical scatterer, both the evanescent-field ratio $E_x/E_z$ and the effective polarizability are invariant under translations within the homogeneous cladding as shown in Fig.~\ref{fig:2.1.1}(a), provided that the scatterer remains sufficiently far from the interface for boundary-induced corrections to its polarizability to be negligible. The leading electric-dipole backscattering is therefore suppressed independently of its position within this region. For a fixed raised-bump geometry, translational invariance of the unperturbed planar interface likewise preserves both the local modal polarization and the effective polarizability, yielding the same leading-order cancellation under translations of the scatterer along the interface as shown in Fig.~\ref{fig:2.1.1}(c).

The same mechanism can be extended beyond isolated Type-2 scatterers to fabrication-induced sidewall or surface roughness, which is a common source of scattering loss in dielectric waveguides. We consider a slowly varying rough interface described by the local normal displacement $\Delta h(x)$ relative to the unperturbed flat core--cladding interface, where $x$ denotes the coordinate along the interface. Within the locally planar shifted-boundary approximation \cite{johnsonPerturbationTheoryMaxwells2002a,johnson2005roughness,kitaAreSlotSubwavelength2018}, the leading-order response of the rough interface can be represented as a continuous distribution of local effective dipolar responses characterized by a position-dependent diagonal effective polarizability tensor $\boldsymbol{\alpha}_{\mathrm{eff}}(x)$. Its components are
\begin{equation*}
    \alpha_{xx}^{\mathrm{eff}}(x)=\Delta h(x)\Delta\varepsilon,
    \qquad
    \alpha_{zz}^{\mathrm{eff}}(x)=-\Delta h(x)\varepsilon_{\mathrm{cl}}^2\Delta(\varepsilon^{-1}),
\end{equation*}
where $\Delta\varepsilon=\varepsilon_{\mathrm{co}}-\varepsilon_{\mathrm{cl}}$ and $\Delta(\varepsilon^{-1})=\varepsilon_{\mathrm{co}}^{-1}-\varepsilon_{\mathrm{cl}}^{-1}$. Using $\rho=\varepsilon_{\mathrm{co}}/\varepsilon_{\mathrm{cl}}$, their ratio becomes
\begin{equation*}
    \frac{\alpha_{zz}^{\mathrm{eff}}(x)}
         {\alpha_{xx}^{\mathrm{eff}}(x)}
    =
    \frac{1}{\rho},
\end{equation*}
which is independent of the local boundary displacement $\Delta h(x)$. The polarization-matching condition in Eq.~(\ref{eq:3.5}) can therefore suppress the leading-order backscattering contribution from this spatially distributed class of boundary disorder. Kita \textit{et al.} showed that the locally flat shifted-boundary approximation is already accurate for Gaussian-correlated roughness when the correlation length $L_c$ is approximately an order of magnitude larger than the RMS roughness amplitude $\sigma$, i.e., $L_c/\sigma\gtrsim10$ \cite{kitaAreSlotSubwavelength2018}. Experimentally measured sidewall roughness in state-of-the-art SOI waveguides typically satisfies this condition, making this regime directly relevant to practical fabrication-induced disorder.

\subsection{Circular dichroism engineering}

The cancellation condition in Eq.~(\ref{eq:3.7}) can be approached over an extended region of a lossless isotropic dielectric waveguide by introducing circular dichroism to the waveguide through structural engineering. One way to realize such circular dichroism is to open a polarization-selective band gap in a waveguide supporting LCP-like and RCP-like guided modes, such as a circular fiber, by introducing a chiral structural perturbation that selectively Bragg-couples the counter-propagating modes of one circular-polarization handedness while leaving the opposite handedness propagating \cite{koppDoublehelixChiralFibers2003,leePolarizationGapsSpiral2005,chenObservationBackscatteringImmuneChiral2011,takahashiCircularDichroismThreedimensional2014,takahashiOpticalPropertiesChiral2023}.

\begin{figure}[h]
    \centering
    \includegraphics[width=\linewidth,height=0.6\textheight,keepaspectratio]{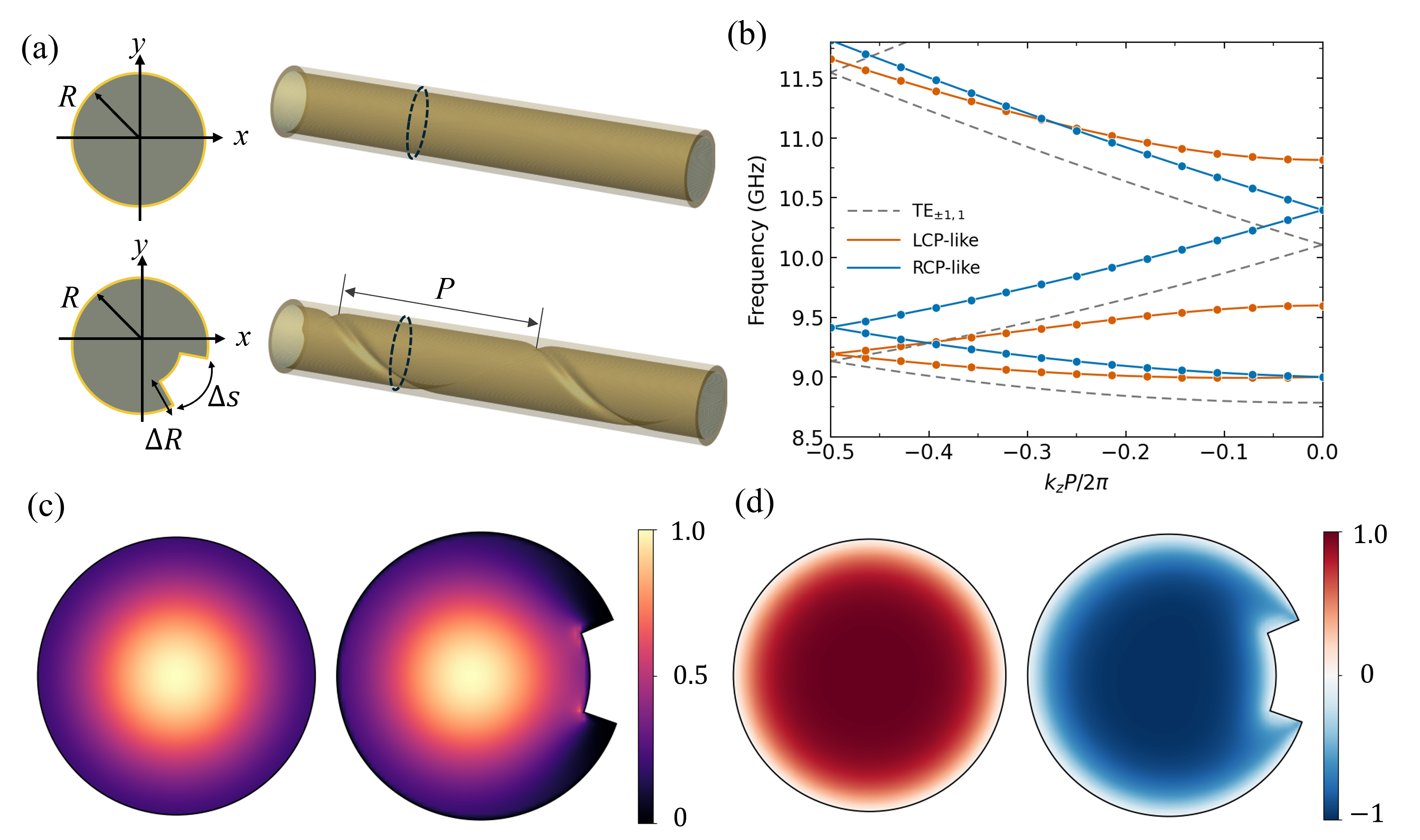}
    \caption{Structural circular dichroism in the chiral PEC fiber. (a) Schematic of the unperturbed straight PEC fiber and the helical PEC fiber. The fiber radius is \(R\), while the boundary protrusion has radial depth \(\Delta R\), arc length \(\Delta s\), and helical pitch \(P\). (b) Calculated band structure of the helical PEC waveguide. The dashed curves denote the degenerate \(\mathrm{TE}_{\pm1,1}\) modes in the unperturbed circular fiber, while the solid curves denote the LCP-like and RCP-like modes in the helically perturbed structure which originate from \(\mathrm{TE}_{1,1}\) and \(\mathrm{TE}_{-1,1}\) respectively. The helical boundary modulation opens a polarization-selective gap at the \(\Gamma\) point shaded in blue. (c) Normalized electric-field intensity of the forward-moving \(\mathrm{TE}_{1,1}\) mode in the straight PEC fiber and RCP-like mode helical PEC fibers at 10GHz, respectively. (d) The normalized Stokes parameter \(s_3\) for the forward-moving $\mathrm{TE}_{1,1}$ mode of the straight PEC fiber and RCP-like mode in the helical PEC fiber at 10GHz. Both modes retain a high degree of circular polarization over a substantial interior region of the fiber. Their opposite signs of \(s_3\) show that the surviving RCP-like mode has the opposite polarization handedness to the \(\mathrm{TE}_{1,1}\) mode, confirming \(\mathrm{TE}_{1,1}\)-derived branch is the one gapped in in (b).}
    \label{fig:2.2.1}
\end{figure}

To illustrate how such structural circular dichroism can be engineered to realize the cancellation condition in Eq.~(\ref{eq:3.7}), we redesigned the helical perfect-electric-conductor (PEC) waveguide introduced in \cite{chengIncompleteImmunityBackscattering2015a}. As schematically shown in Fig.~\ref{fig:2.2.1}(a), we consider an unperturbed air-filled circular PEC fiber of radius \(R\) and a helically perturbed counterpart formed by introducing a helical groove along the PEC boundary. The groove is characterized by a radial depth \(\Delta R\), an outer arc length \(\Delta S\), and a helical pitch \(P\). We define the groove as right-handed when its azimuthal position \(\phi\) increases with \(z\), and adopt this convention throughout. In the unperturbed circular PEC fiber, the fundamental modes are \(\mathrm{TE}_{m,1}\) with azimuthal order \(m=\pm1\). The forward- and backward-propagating fields of these two fundamental modes can be written as
\begin{equation}
    \begin{aligned}
        \mathbf{E}_{+k}^{(m)}(\mathbf{r}) & =e^{i(m \phi+k z)}[A(\rho) \hat{\boldsymbol{\rho}}+i m B(\rho) \hat{\boldsymbol{\phi}}], \\
        \mathbf{E}_{-k}^{(m)}(\mathbf{r}) & =e^{i(m \phi-k z)}[A(\rho) \hat{\boldsymbol{\rho}}+i m B(\rho) \hat{\boldsymbol{\phi}}], \quad m= \pm 1,
    \end{aligned} \label{eq:3.2.1}
\end{equation}
where \((\rho,\phi,z)\) are cylindrical coordinates with \(z\) along the waveguide axis, \(k>0\) is the magnitude of the longitudinal propagation constant, and \(\hat{\boldsymbol\rho}\) and \(\hat{\boldsymbol\phi}\) are the radial and azimuthal unit vectors. The real radial functions \(A(\rho)\) and \(B(\rho)\) are proportional to \(J_1(\gamma\rho)/(\gamma\rho)\) and \(J_1'(\gamma\rho)\), respectively, where \(J_1\) is the first-order Bessel function, \(J_1'\) is its derivative, and \(\gamma\) is the transverse wave number of the \(\mathrm{TE}_{11}\) mode. Those fundamental $\mathrm{TE}$ modes in the unperturbed PEC fiber has zero $E_z$ components, and the radial and azimuthal field components are in quadrature but generally have unequal amplitudes, so the local polarization is elliptical. As \(\rho\rightarrow0\), \(A(\rho)\rightarrow B(\rho)\), and the \(\mathrm{TE}_{+1,1}\) and \(\mathrm{TE}_{-1,1}\) modes approach opposite perfect circular polarizations. We therefore adopt the optics convention to label those fundamental modes as LCP-like or RCP-like according to the handedness of their transverse-field rotation relative to the propagation direction. The forward \(\mathrm{TE}_{+1,1}\) and backward \(\mathrm{TE}_{-1,1}\) modes constitute the LCP-like time-reversal pair, whereas the forward \(\mathrm{TE}_{-1,1}\) and backward \(\mathrm{TE}_{+1,1}\) modes constitute the RCP-like pair.

The helical boundary modulation can be expanded in helical spatial harmonics as
\begin{equation}
    \Delta \varepsilon_c(\rho, \phi, z)=\sum_l \Delta \varepsilon_c^{(l)}(\rho) e^{i l(\phi-q z)}, \quad q=\frac{2 \pi}{P} \label{eq:3.2.2}
\end{equation}
where $P$ is the helical pitch. The right-handed modulation selectively couples LCP-like pair of fundamental modes, namely the forward \(\mathrm{TE}_{-1,1}\) mode at $\mathrm{k}=q$ to its time-reversal counterpart \(\mathrm{TE}_{+1,1}\) at $\mathrm{k}=-q$, because the \(l=2\) harmonic of the modulation as seen in Eq.~(\ref{eq:3.2.2}) supplies both the required change in azimuthal order from \(-1\) to \(+1\) and the longitudinal wave-number change \(2q\). The corresponding unperturbed states at $\mathrm{k}=\pm q$ are folded to the $\Gamma$ point of the reduced Brillouin zone, where their coupling will open a polarization-selective gap. 

\begin{figure}[h]
    \centering
    \includegraphics[width=\linewidth,height=0.6\textheight,keepaspectratio]{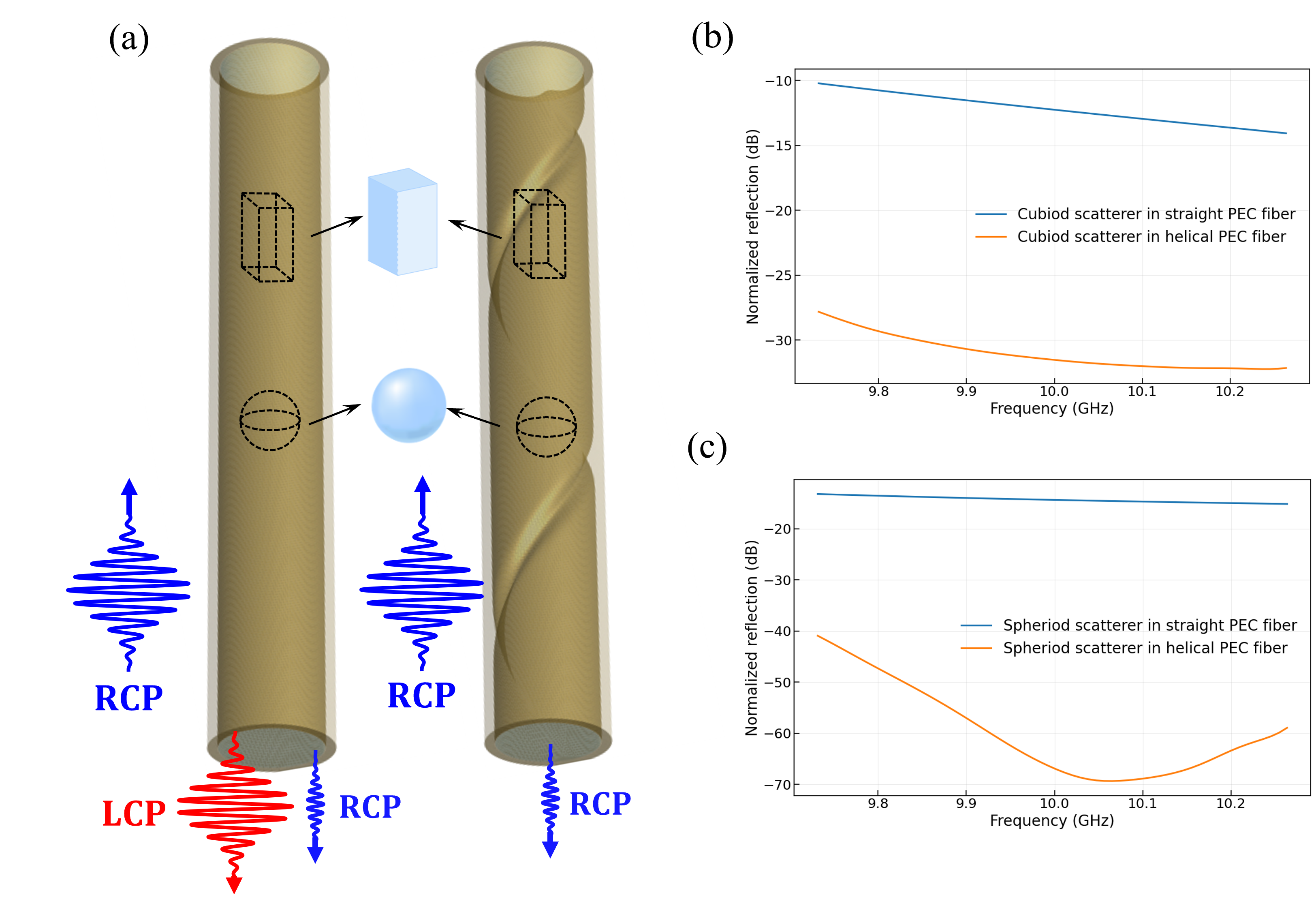}
    \caption{Suppression of disorder-induced backscattering in the helical PEC fiber.(a) Schematic of the scattering simulations in the straight and helical PEC fibers. The \(+z\)-propagating RCP-like guided mode serves as the background incident mode. A cuboid or spheroidal dielectric scatterer is placed on the fiber axis, with the two geometries simulated separately. In the straight fiber, the dominant first-Born backscattering channel is the \(-z\)-propagating LCP-like mode. In the helical fiber, this channel is removed from the propagating spectrum by the polarization-selective gap, leaving only a much weaker RCP-like reflected component arising from higher-order Born contributions. (b) Normalized reflection spectra for the cuboid scatterer in the straight and helical PEC fibers. (c) Normalized reflection spectra for the spheroidal scatterer. In both (b) and (c), the helical PEC fiber strongly suppresses reflection relative to the straight fiber, consistent with the polarization-induced cancellation condition in Eq.~(\ref{eq:3.7}).}
    \label{fig:2.2.2}
\end{figure}

We verify the polarization-selective gap and how the cancellation condition in Eq.~(\ref{eq:3.7}) can suppress the backscattering induced by the Type-I scatterer regardless their geometry shape by full-wave FDTD simulations. For the detailed parameters, we take the radius of the unperturbed air-filled circular PEC waveguide to be \(R=1~\mathrm{cm}\). The helical groove on the boundary is a truncated PEC sector with radial depth \(\Delta R=0.25R\) and outer arc length \(\Delta S=0.75R\), and the helical pitch is chosen as \(P=6~\mathrm{cm}\). The calculated band structure in Fig.~\ref{fig:2.2.1}(b) exhibits a broad polarization-selective gap, shaded in blue. Within this frequency range, the LCP-like branch derived from the \(\mathrm{TE}_{1,1}\) sector is gapped, whereas the RCP-like time-reversal pair, comprising the forward \(\mathrm{TE}_{-1,1}\) and backward \(\mathrm{TE}_{+1,1}\) modes, remains propagating. Figure~\ref{fig:2.2.1}(c) compares the normalized electric-field intensities of the forward \(\mathrm{TE}_{1,1}\) mode in the straight PEC fiber and the surviving forward RCP-like mode in the helical PEC fiber at \(10~\mathrm{GHz}\). Their corresponding polarization structures are characterized in Fig.~\ref{fig:2.2.1}(d) by the normalized Stokes parameter \(S_3(\mathbf{r})\) \cite{born1999principles}:
\begin{equation}
    S_3(\mathbf{r})=\frac{2 \operatorname{Im}\left[E_{\rho}(\mathbf{r}) E_{\phi}^*(\mathbf{r})\right]}{\left|{E_{\rho}}(\mathbf{r})\right|^2+\left|E_{\phi}(\mathbf{r})\right|^2}
\end{equation}
where \(|S_3|\) quantifies the local degree of circular polarization, while its sign specifies the polarization handedness \cite{born1999principles}. Although perfect circular polarization is theoretically only attained in the \(\rho\rightarrow0\) limit for \(\mathrm{TE}_{1,1}\) in the straight PEC fiber, the numerical results in Fig.~\ref{fig:2.2.1}(d) show that a high degree of circular polarization extends over a substantial interior region away from the PEC boundary in the straight fiber and remains largely preserved in the helical fiber. The opposite signs of \(S_3\) for the straight-fiber \(\mathrm{TE}_{1,1}\) mode and the surviving RCP-like mode further confirm their opposite polarization handedness, consistent with the \(\mathrm{TE}_{1,1}\) branch being removed by the polarization-selective gap. We next examine the suppression of scatterer-induced reflection by full-wave simulations. As illustrated in Fig.~\ref{fig:2.2.2}(a), dielectric scatterers with relative permittivity \(\varepsilon_{\mathrm{sca}}=2\) are placed on the fiber axis, where the forward \(\mathrm{TE}_{-1,1}\) background mode, \(\mathbf{E}_{+k}^{(-1)}(\mathbf r)\), is nearly perfect circular-polarized. Each scatterer has a characteristic transverse dimension of \(0.5R\) and a longitudinal extent of \(0.5P\). A cuboid and a spheroidal scatterer are considered separately, and for each geometry an identical scatterer is placed at the corresponding position in the unperturbed straight PEC fiber for comparison. In the straight fiber, the dominant first-Born backscattering channel is the \(-z\)-propagating LCP-like mode. In the helical fiber, this channel is removed from the propagating spectrum by the polarization-selective gap, leaving only a much weaker RCP-like reflected component arising from higher-order scattering. Figure~\ref{fig:2.2.2}(b) compares the normalized reflection spectra for the cuboid scatterer and shows an approximately \(20~\mathrm{dB}\) reduction throughout the operating frequency range in the helical fiber relative to the straight fiber. A comparable suppression is obtained for the spheroidal scatterer, as shown in Fig.~\ref{fig:2.2.2}(c), despite its substantially different geometry. TThese results demonstrate that the suppression of first-Born backscattering by structural circular dichroism is not restricted to a particular scatterer geometry. For dielectric perturbations dominated by first-Born backscattering, this mechanism can therefore substantially reduce disorder-induced reflection in a reciprocal isotropic waveguide.

The resulting suppression should not be interpreted as completely backscattering immunity. For a practical scatterer or disorder with finite dielectric contrast and finite spatial extent, it would induces higher-order Born backscattering that are not removed by Eq.~(\ref{eq:3.7}). Nevertheless, for the broad class of perturbations for which first-Born backscattering provides the dominant contribution, structural circular dichroism provides an effective mechanism for substantially reducing reflection in a reciprocal isotropic waveguide. In addition, the fundamental mode of the periodically perturbed waveguide is only approximately circularly polarized away from the waveguide axis, so Eq.~(\ref{eq:3.7}) is not satisfied exactly throughout a scatterer of finite spatial extent. This limitation is specific to the present implementation rather than to the cancellation mechanism of Eq.~(\ref{eq:3.7}) itself \cite{SupplementaryMaterial}. The residual reflection observed in Fig.~\ref{fig:2.2.2} can therefore arise from higher-order scattering, deviations from perfect circular polarization over the scatterer volume, and numerical errors associated with the FDTD discretization \cite{SupplementaryMaterial}.

\section{Analysis of photonic topological insulator model}
We now examine the implications of the general argument above for representative reciprocal photonic topological insulators (PTIs). The extent to which reciprocal photonic topological insulators (PTIs) can, in general, provide genuine protection against disorder-induced backscattering has received increasing scrutiny in recent years \cite{orazbayev2019quantitative,rosiek2023observation,khanikaev2024topological}. We argue that the reciprocity fundamentally prevents these systems from providing universal protection against backscattering: for any reciprocal PTI waveguide, there always exists a physically admissible reciprocal disorder that couples the counter-propagating channels and induces back-reflection. In this section, we illustrate this limitation using representative valley-Hall and spin-Hall PTI waveguides, and further demonstrate through full-wave simulations that suitably chosen localized reciprocal scatterers can induce substantial reflection in both systems.

\subsection{Valley-Hall PTI}
The valley-Hall PTI emulates valley-Hall (VH) effect in the condensed matter physics to construct the topological protected edge states. The valley-Hall effect arises from broken inversion symmetry in a 2D honeycomb lattice that leads to opposite Berry curvatures peak at the two valleys $K$ and $K'$ in the band structure. The integration of the local berry curvature around $K(K')$ gives rise to “valley Chern index”, denoted by $C_{K}$($C_{K'}$). A typical valley hall photonic topological insulator (VH-PTI) waveguide is formed by interfacing two valley hall photonic crystal (VPC) lattices characterized by opposite half-integer valley Chern indices, resulting in an integer valley Chern indices difference: $\Delta C_{K/K'} = \pm1$ across the interface. The edge states appearing on the interface with certain shape (eg. bridge-shaped and zigzag-shaped) could be used to guide the light along the interfacing direction. It is claimed that those edge states, which is dependent on the shape of edge, are topologically protected by the difference of valley Chern number $\Delta C_{K/K'} = \pm1$ between the two bulks \cite{noh2018observation}\cite{ma2016allsi} so they could provide certain robustness against the imperfections and sharp corners in the integrated photonic devices. 

\begin{figure}[b]
    \centering
    \includegraphics[width=\linewidth]{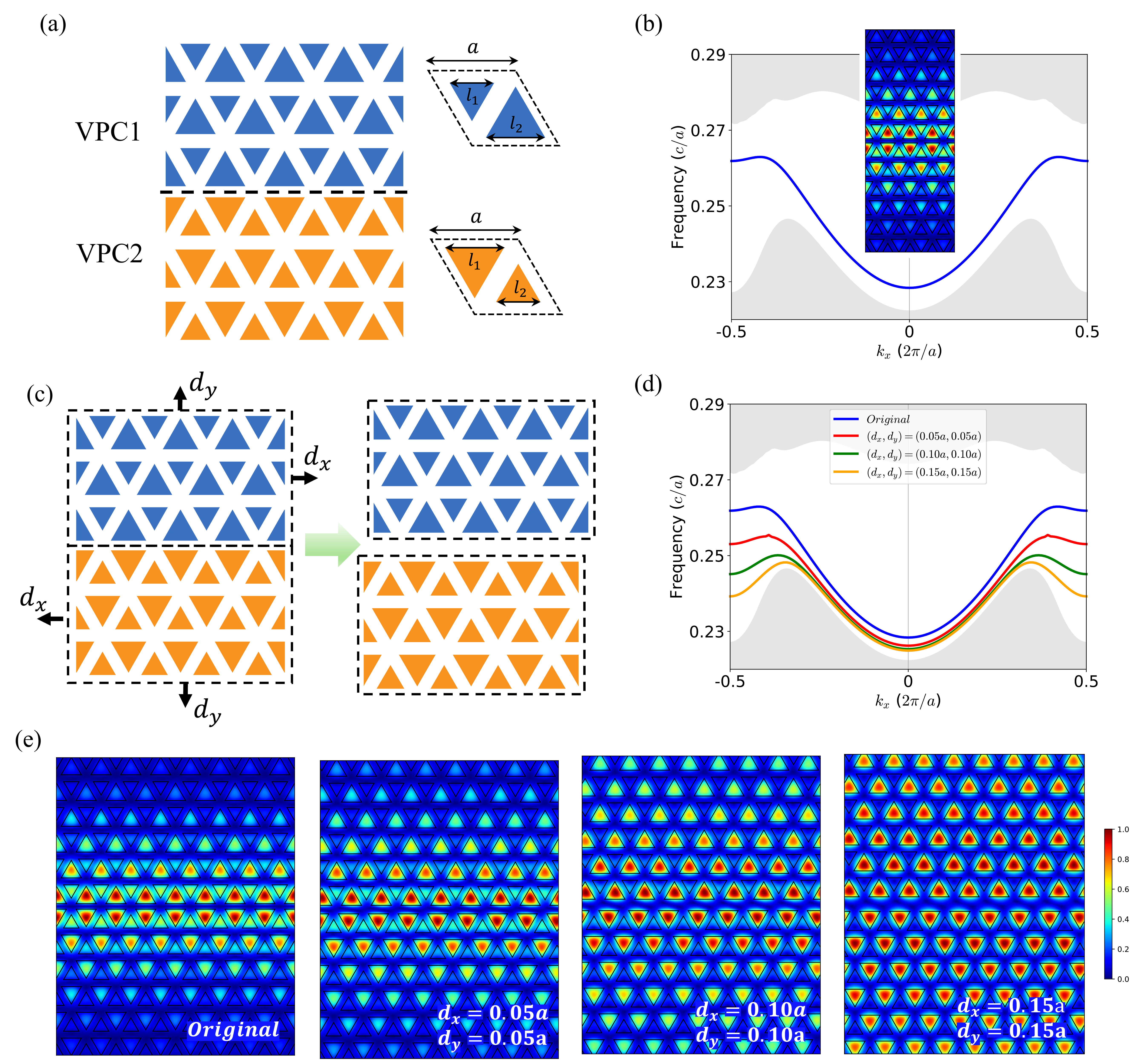}
    \caption{Valleyhall PTI structure and its band diagram. (a) Schematics of the unit cell of two topological distinct VPC with the pointing-up and pointing down triangles embeded within the air background. (b) The structure of the valleyhall PTI, where the dashed line denotes the interface formed by two VPCs. The blue and red arrows represents the forward-moving and backward-moving edge mode locked to different valleys. The shaded area is the periodic structure of the wavefunction of the edge modes along the direction of the interface. (c) The dispersion relation edge mode. The red and blue line represents the dispersion of the edge mode locked to different valleys. The forward-moving edge mode denoted by the blue dot can be coupled to the backward-moving edge mode denoted by the red dot by local defect due to the reciprocity}
    \label{fig:2.1}
\end{figure} 

Unfortunately, we argue that such dielectric VH-PTI waveguides fall significantly short of providing universal protection against disorder-induced backscattering. This is a direct consequence of the reciprocity-based argument established above: the forward- and backward-propagating valley edge states remain a reciprocal pair, and the valley degree of freedom does not provide a universal selection rule that forbids their coupling under reciprocal perturbations. Many studies argue that VH-PTI waveguides can suppress reflection provided that the disorder does not induce intervalley scattering. While this condition can indeed lead to reduced backscattering for restricted classes of disorder, a generic localized disorder can provide the large momentum transfer required to couple the \(K\) and \(K'\) valleys and thereby induce back-reflection. In particular, a perturbation localized within one or a few unit cells generally possesses a broad spatial-frequency spectrum capable of overlapping the momentum separation between the two valley-polarized edge states. Consequently, one can always find a physically admissible reciprocal local disorder that couples the counter-propagating valley channels and produces back-reflection.

Furthermore, the ability of VH PTI waveguide to suppress intra-valley scattering \cite{ma2016allsi}\cite{noh2018observation}\cite{yang2020terahertz}\cite{guo2024quantifying} should not be boasted as a distinctive advantage of VH PTI over topological trivial photonics system as such suppression can be readily achieved in a conventional PhC waveguide with sloped-band dispersion \cite{dai2025high}. Finally, the inability of armchair type VH PTI to support valley-dependent edge states is symptomatic of an underlying problem of valley-Hall PTI model: the valley Chern number only characterizes local topological property while so far no bulk-edge correspondence theory has established that such “local” topological index sufficiently guarantees the topological edge mode. It is therefore anticipated that by changing the configuration of the edge in VH PTI waveguide (eg. shifting the two bulks away along vertical/horizontal direction) one could alter the dispersion of the edge mode and make it unfavorable for light transport. In summary, VH PTI waveguide theoretically offers no superiority than the conventional band gap waveguide. 
\begin{figure}
    \centering
    \includegraphics[width=\linewidth,height=0.6\textheight,keepaspectratio]{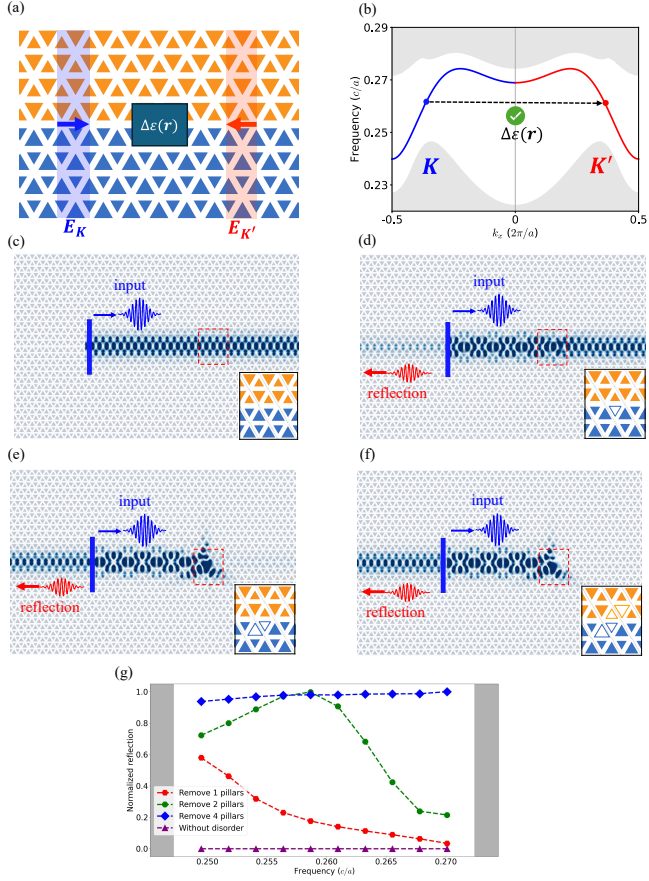}
    \caption{Coupling of valley-polarized edge states by localized reciprocal disorder in a VH photonic-crystal waveguide. (a) Schematic of the interface with a localized permittivity perturbation \(\Delta\varepsilon(\mathbf r)\) and the counter-propagating edge states associated with the \(K\) and \(K'\) valleys. (b) Edge-state dispersion, with the forward- and backward-propagating branches associated predominantly with the \(K\) and \(K'\) valleys, respectively. (c) Out-of-plane electric-field distribution \(E_z\) in the disorder-free waveguide. (d)--(f) Corresponding field distributions after removing one, two, and four pillars near the interface, respectively. Pronounced backward-propagating fields emerge as the local disorder is introduced. (g) Normalized reflection spectra for the disorder-free waveguide and the three defect configurations. Removing four pillars produces nearly unity reflection over a broad portion of the edge-state frequency range. The gray regions indicate the projected bulk bands.}
    \label{fig:2.2}
\end{figure} 

To demonstrate our point, we construct a 2D valleyhall PhC waveguide as seen in Fig following the design scheme in \cite{ma2016allsi}\cite{chen2017valley}\cite{yang2020terahertz}. The valleyhall PTI is formed by connecting two honeycomb photonic crsytal (VPC1 and VPC2) with a bridge-shape interface denoted by the dashed line in Fig.\ref{fig:2.1}(a). The unit cell of VPC1 (blue) and VPC2 (orange) comprises two inverted equilateral triangular pillars with the relative permitivty $\varepsilon=11.7$ embeded in the air background. For VPC1, the size parameter of the upward-pointing rod ($\triangle$) is $l_1 = 0.68a$, while that of the downward-pointing rod ($\triangledown$) is $l_2 = 0.51a$ (blue rhombic cell in the inset of Fig.\ref{fig:2.1}(a), where $a$ denotes the lattice constant. In contrast, VPC2 adopts the inverse configuration, with the size parameters interchanged: $l_2$ for the $\triangle$ rod and $l_1$ for the $\triangledown$ rod (orange rhombic cell in the inset of Fig.\ref{fig:2.1}(a). The detailed photonic band structure and non-trivial topology of the designed VPC is shown in the supplementary. We focus on the transverse-magnetic (TM) modes with non-zero out-of-plane $E_z$ and in-plane $H_x$ and $H_z$ componenets, and other components being zero. The dispersion of the edge state that exists at the interface between two VPCs is shown in Fig.\ref{fig:2.1}(b) with the inset shows the $E_z$ field intensity distribution of the edge state at the projected $K$ valley. It is claimed that the dispersion of the edge states on such a bridge-shape edge results from the difference between the valley Chern indices of VPC1 and VPC2. However, the gapped edge dispersion in Fig.~\ref{fig:2.1}(b) already confirms the system is topological trivial. The most important property of the topological edge state is that the gapless edge states persist as long as the bulk band gap is not closed. In the VH PTI if we slightly move the VPC1 and VPC2 in the opposite direction with the displacement $d_x$ and $d_y$ in the $x$ and $y$ direction respectively as illustrated in Fig.\ref{fig:2.1}(c), we could see from Fig.\ref{fig:2.1}(d-f) that the dipersion of the edge state gradually move downward and merge with the bulk states below the band gap. It is worth to mention that such variations do not change the valley Chern number, do not induce inter-valley coupling, and most importantly, do not close the bulk band gap. The dissolving of the edge states could be more intuitively visualized from the insets Fig we could see from Fig, as we move the VPC1 and VPC2 apart. This indicates that the existence of the counter-propagating edge mode in VH PTI is not guaranteed by the valley Chern number, and the true sufficient condition of the propagating edge mode in this type of valleyhall PhC waveguide is worth to explore.
\begin{figure}[h]
    \centering
    \includegraphics[width=\linewidth,height=0.6\textheight,keepaspectratio]{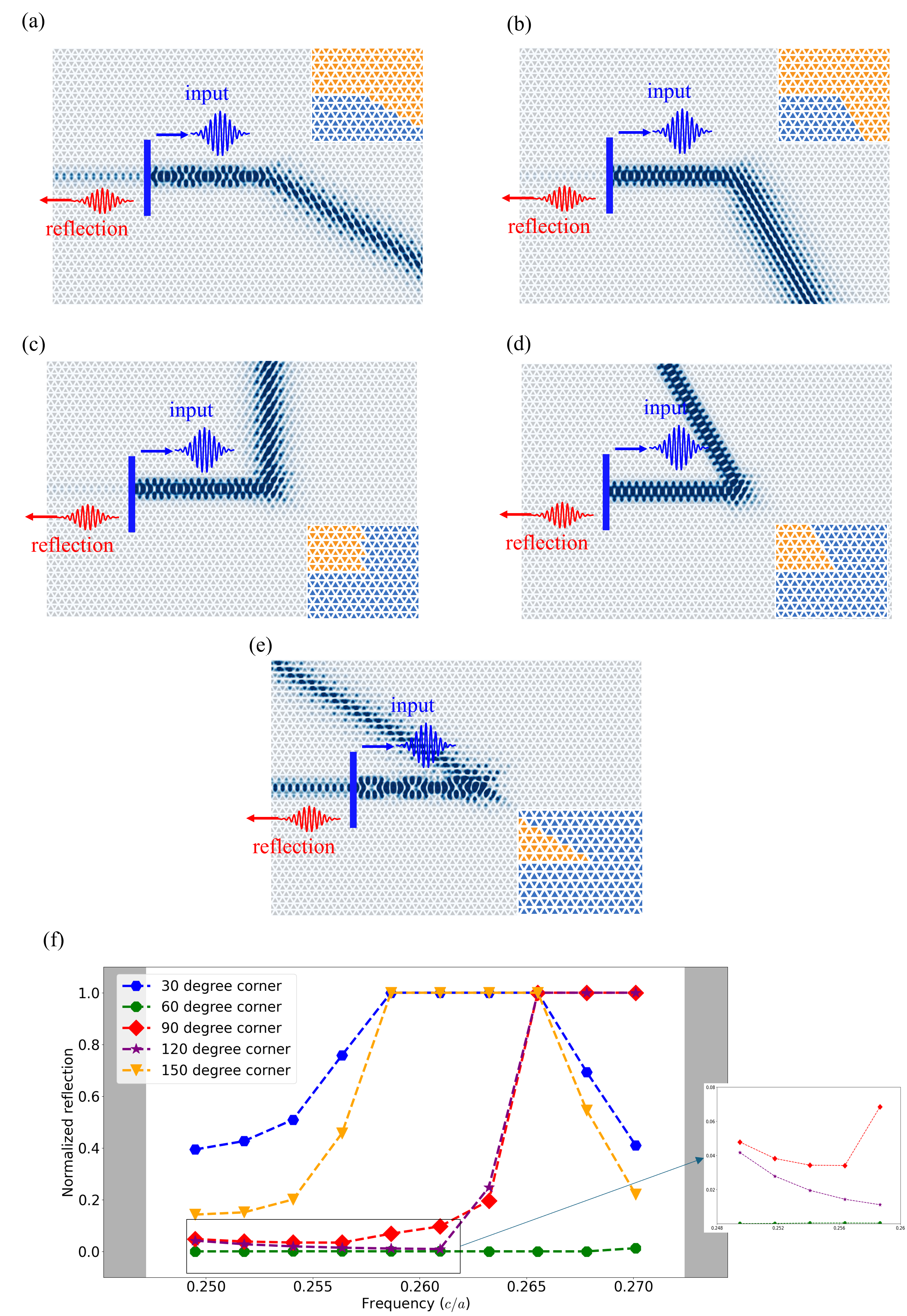}
    \caption{Simulation result of Valley Hall waveguide with sharp bending of different bending angles. (a-e) The out-of-plane $\mathbf{E}_z$ intensity distribution of the Valley Hall PhC waveguide with the sharp bending of 30 degrees (a), 60 degrees (b), 90 degrees (c), 120 degrees (d), 150 degrees (e). (g) The reflection spectra measured at different frequency points within the band gap of VPCs caused by removing different number of the pillars from the edge. The shaded regions at the both ends of spectra are the bulk states. The dashed line connecting the measured points is used for guiding the sight}
    \label{fig:2.3}
\end{figure} 

For the bridge-shaped interface, another commonly studied configuration is obtained by interchanging VPC1 and VPC2, such that the large triangular pillars rather than the small pillars face each other across the interface, as shown in Fig.~\ref{fig:2.2}(a). We use this configuration to explicitly demonstrate the coupling of the counter-propagating valley edge states by localized reciprocal disorder. The corresponding edge-state dispersion is shown in Fig.~\ref{fig:2.2}(b), where the forward- and backward-propagating modes are associated predominantly with the \(K\) and \(K'\) valleys, respectively. As discussed above, valley separation does not constitute a universal selection rule against reciprocal perturbations: a sufficiently localized structural perturbation contains the spatial-frequency components required to couple the two valley sectors.

We introduce a simple artificial disorder by removing one, two, or four pillars near the interface. Such pillar removal modifies only the local dielectric structure and preserves reciprocity, while strongly breaking the translational structure responsible for the approximate separation of the two valley channels. In the absence of disorder, Fig.~\ref{fig:2.2}(c) shows essentially unidirectional transmission along the interface with negligible backward field. Once the localized defects are introduced, pronounced reflected waves appear in Figs.~\ref{fig:2.2}(d)--(f). The corresponding reflection spectra in Fig.~\ref{fig:2.2}(g) show that even a small number of removed pillars can generate substantial and strongly frequency-dependent back-reflection. Most notably, removing four pillars produces reflection close to unity over a broad portion of the edge-state frequency range, nearly suppressing forward transmission. These full-wave results provide a direct illustration of the reciprocity-based argument above: although intervalley scattering may be weak for restricted classes of smooth disorder, physically admissible localized reciprocal perturbations can couple the counter-propagating valley edge states and induce substantial back-reflection.

We further investigate the response of the VH-PTI waveguide to sharp bends. We emphasize that a sharp bend is not equivalent to the localized disorder considered in our general scattering argument, since it represents a finite change of the waveguide geometry rather than a compact perturbation of an otherwise fixed interface. Nevertheless, reciprocity again leaves a counter-propagating channel available, so the valley-Hall topology does not provide universal immunity against reflection from arbitrary bending geometries. Figure~\ref{fig:2.3} compares bends of \(30^\circ\), \(60^\circ\), \(90^\circ\), \(120^\circ\), and \(150^\circ\). While some configurations, most notably the \(60^\circ\) and \(120^\circ\) bends, exhibit very low reflection over part of the operating range, other bending geometries produce strong or nearly complete reflection. In particular, the unusually high transmission of the \(120^\circ\) bend has recently been shown to depend strongly on the detailed interface geometry rather than on the valley-Hall topology \cite{dai2025high}. Our results thus reveal the same general limitation from a complementary geometric perspective: reciprocity leaves a backward-propagating channel available, and although certain bend geometries can strongly suppress coupling to this channel, the resulting low reflection is not universal.

\subsection{Spin-Hall PTI}
Photonic analogues of the quantum spin-Hall (QSH) effect provide another important class of reciprocal PTIs, in which additional symmetries are engineered to construct and protect photonic pseudospin sectors \cite{wu2015scheme,khanikaev2013photonic}. Here, we focus on the widely studied class in which the pseudospins are constructed from crystalline symmetry. \cite{wu2015scheme,khanikaev2013photonic}. Unlike the Kramers protection of electronic QSH systems, however, the decoupling of these photonic pseudospins is not enforced by reciprocity itself. A generic reciprocal perturbation need not preserve the additional crystalline symmetry and can therefore mix the counter-propagating pseudospin channels. Consistent with our general argument above, one can thus always find a physically admissible reciprocal disorder that couples the two channels and induces back-reflection.

\begin{figure}[t]
  \centering
  \includegraphics[width=\linewidth]{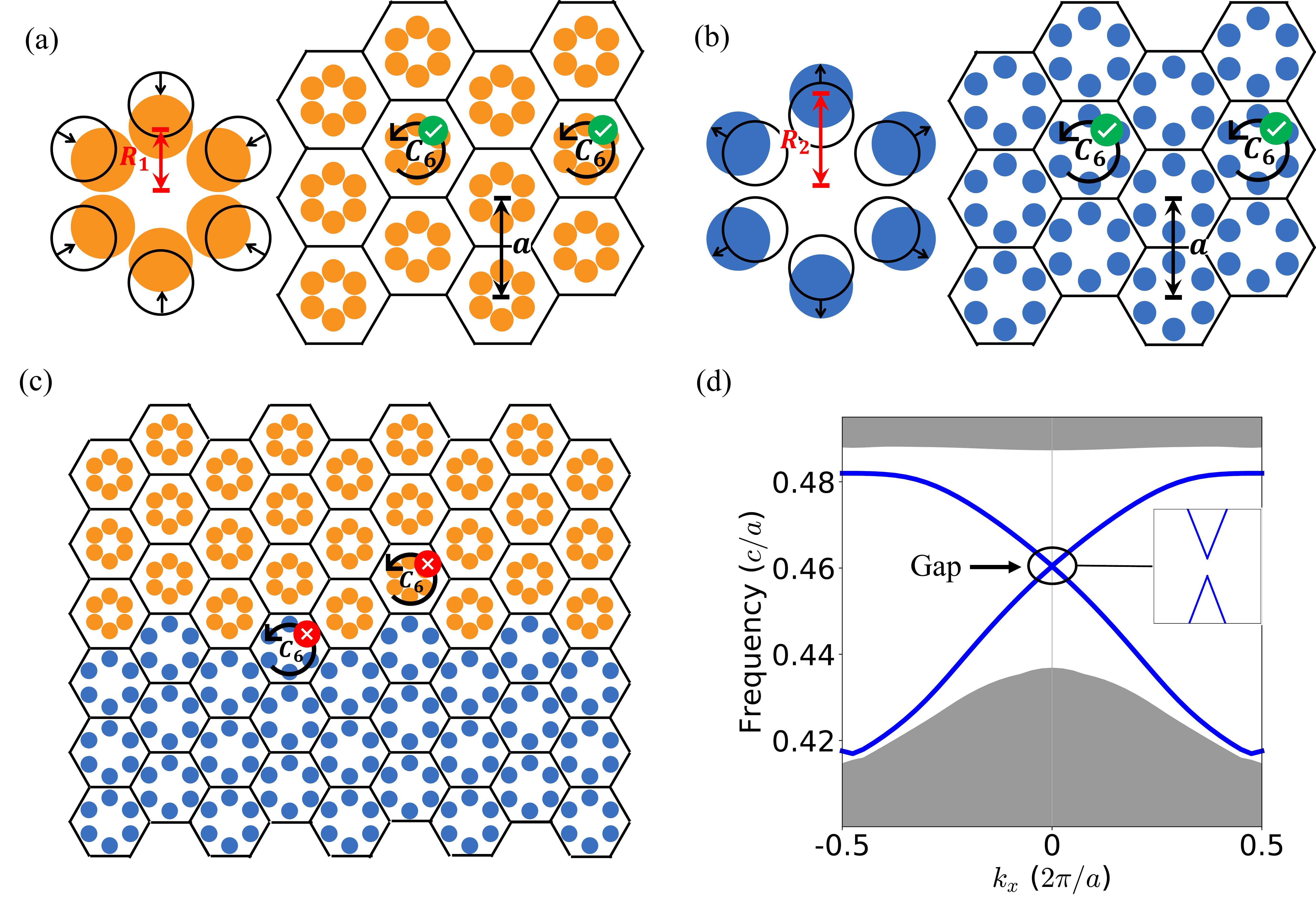}
  \caption{Crystalline-symmetry-based spin-Hall photonic topological waveguide.
    (a,b) Unit cells and bulk lattices formed by two opposite radial displacements of the six dielectric pillars. The \(C_6\) symmetry is preserved in both bulk structures, allowing the construction of the two pseudospin sectors from the corresponding two-dimensional representations. The arrows indicate the displacement of the pillars relative to the reference configuration.
    (c) Interface between the two bulk lattices forming the spin-Hall-like waveguide. The \(C_6\) symmetry is broken for the unit cells intersected by the interface, as indicated by the red crosses.
    (d) Projected edge-state dispersion of the interface. The breaking of the crystalline symmetry at the boundary allows hybridization of the two counter-propagating pseudospin edge states and opens a minigap around \(k_x=0\). The gray regions denote the projected bulk bands.}
  \label{fig:3.1}
\end{figure}

We demonstrate our point following the well-studied design scheme proposed in \cite{wu2015scheme,barik2016two,barik2018topological,parappurath2020direct}. The structure consists of a triangular lattice with a hexagonal unit cell composed of a cluster of six dielectric pillars. The \(C_6\) symmetry of the bulk supports two two-dimensional representations from which the photonic pseudospin states are constructed. Concentrically expanding and shrinking the pillar clusters opens topological and trivial band gaps, respectively, while preserving the \(C_6\) symmetry, as shown in Fig.~\ref{fig:3.1}(a,b). In our implementation, the corresponding cluster radii are \(R_1=a_0/2.75\) and \(R_2=a_0/3.376\), with a pillar radius of \(0.12a_0\), where \(a_0\) is the lattice constant.

However, once the two bulk structures are interfaced to form the waveguide, as shown in Fig.~\ref{fig:3.1}(c), the complete waveguide no longer possesses the \(C_6\) symmetry of the individual bulks, since the interface itself is not invariant under the \(C_6\) rotation. The symmetry that constructs and protects the two pseudospin sectors in the bulk therefore no longer protects the interface modes. The resulting counter-propagating edge states form an ordinary time-reversal-related pair and are allowed to couple, as directly manifested by the minigap opened in the edge-state dispersion in Fig.~\ref{fig:3.1}(d).

The interface therefore supports a pair of counter-propagating edge states related by time-reversal symmetry, but these states are no longer protected as two distinct pseudospin sectors by the \(C_6\) symmetry of the bulk. In particular, there is no remaining symmetry selection rule that forbids coupling between them. The consequence is stronger than the mere existence of reciprocal disorders that can induce backscattering. Since the complete waveguide no longer possesses the \(C_6\) symmetry that defines the bulk pseudospin sectors, there is no corresponding symmetry-preserving class of disorder for which the coupling between the counter-propagating edge states is forced to vanish.

\begin{figure}[t]
    \centering
    \includegraphics[width=\linewidth]{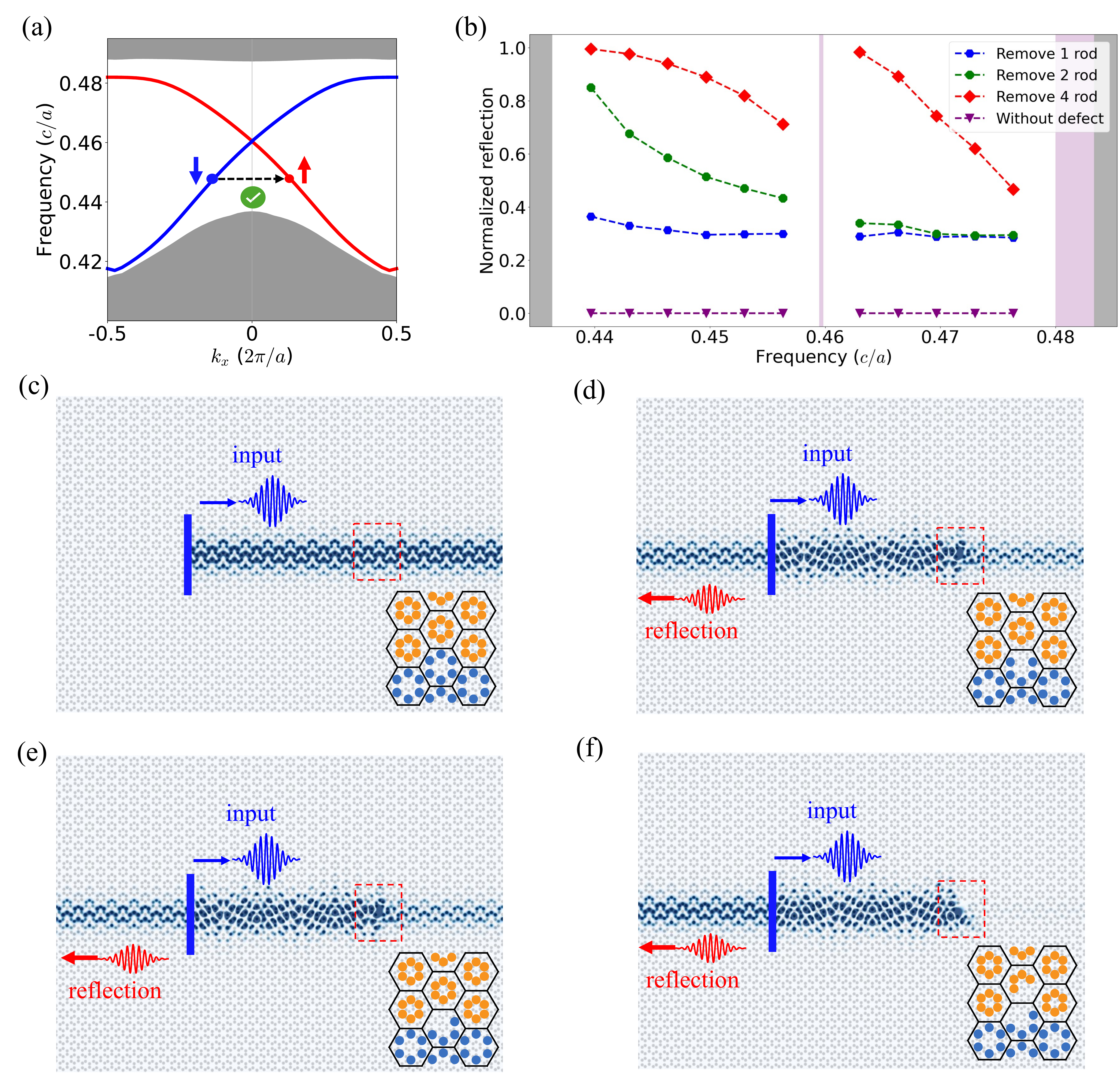}
    \caption{Full-wave simulation of disorder-induced backscattering in the spin-Hall-like topological waveguide. (a) Schematic coupling between the counter-propagating edge states by a localized reciprocal perturbation \(\Delta\varepsilon(\mathbf r)\). The two interface modes are related by time-reversal symmetry but are no longer symmetry-protected as distinct pseudospin sectors once the bulk \(C_6\) symmetry is broken at the interface. (b) Normalized reflection spectra for the disorder-free waveguide and for different pillar-removal defects. (c) Out-of-plane electric-field distribution \(E_z\) in the disorder-free waveguide. (d)--(f) Corresponding field distributions for different numbers of pillars removed near the interface. The insets show the local defect geometries.}
    \label{fig:3.2}
\end{figure}

We demonstrate this directly using full-wave FDTD simulations. An edge mode is injected from the source indicated by the blue line in Fig.~\ref{fig:3.2}(c)--(f), and the backward power flux is measured to the left of the perturbation and normalized to the total power emitted by the source. As a simple class of localized reciprocal disorder, we remove different numbers of dielectric pillars from a unit cell adjacent to the interface. These perturbations modify only the local dielectric geometry and preserve reciprocity. Figure~\ref{fig:3.2}(c) shows the disorder-free reference waveguide, whereas Figs.~\ref{fig:3.2}(d)--(f) show the field distributions in the presence of the local defects at a normalized frequency of \(0.45\). Pronounced backward-propagating fields emerge once the defects are introduced. The corresponding reflection spectra in Fig.~\ref{fig:3.2}(b) further demonstrate that such localized reciprocal defects can induce substantial back-reflection over the edge-state frequency range.

\begin{figure}[!t]
    \centering
    \includegraphics[width=\linewidth,height=0.8\textheight,keepaspectratio]{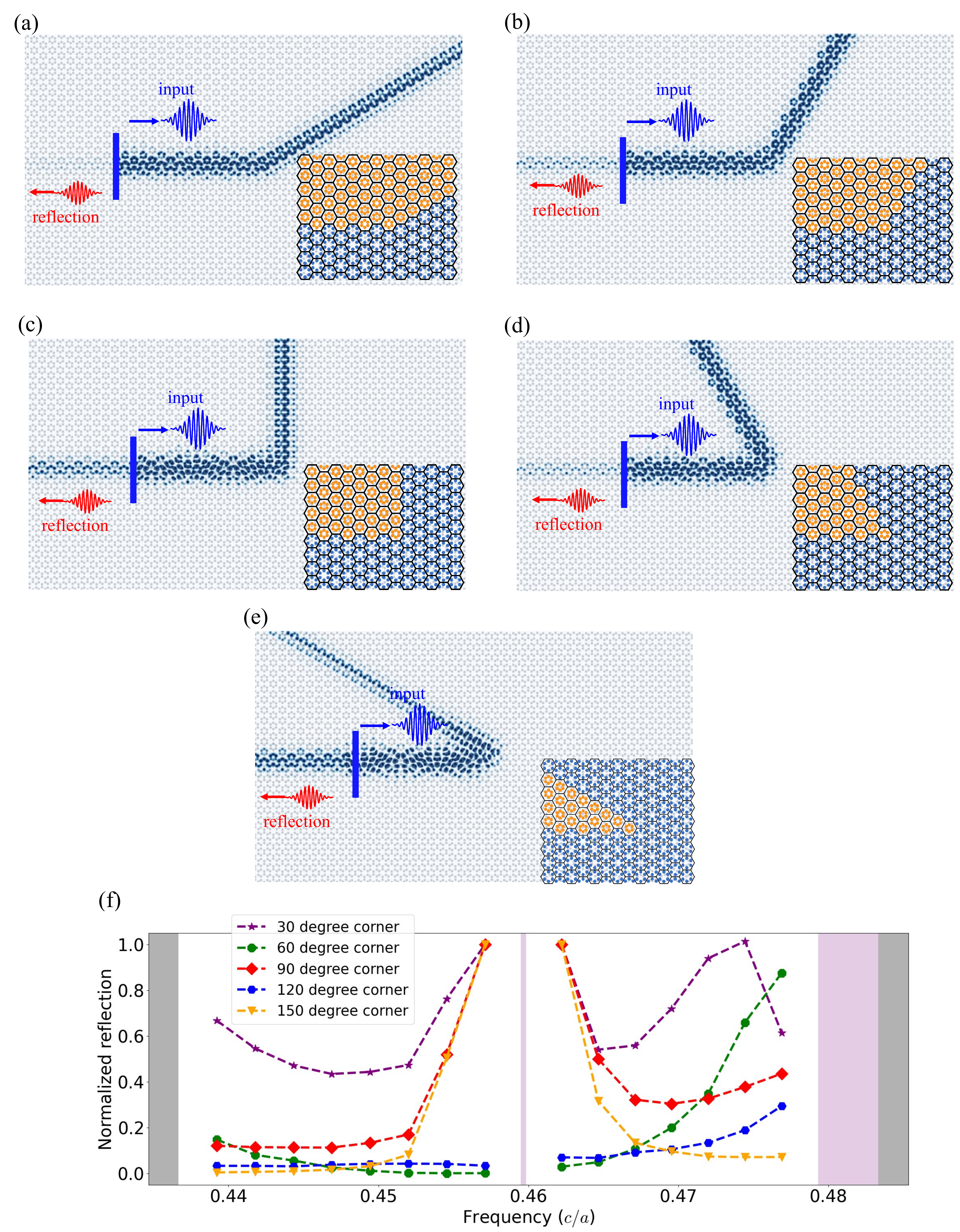}
    \caption{Simulation result of the Spin Hall like topological waveguide with and different sharp bending corner. (a-e) The out-of-plane field $\mathbf{E}_z$ intensity distribution of the waveguide with different sharp bending. The inset shows the configuration of the junction of different bent corner. (f) The spectra of the reflection caused by bent corner on topological waveguide. }
    \label{fig:3.3}
\end{figure} 

We further examine sharp bends with bending angles of \(30^\circ\), \(60^\circ\), \(90^\circ\), \(120^\circ\), and \(150^\circ\), as shown in Fig.~\ref{fig:3.3}. These bends modify the waveguide interface itself and should therefore be distinguished from the localized disorder considered in our general scattering argument. The \(30^\circ\), \(90^\circ\), and \(150^\circ\) bends connect zigzag- and armchair-type interfaces, for which differences in dispersion, field profile, and polarization can produce substantial mode mismatch and strong reflection. By contrast, the \(60^\circ\) and \(120^\circ\) bends provide more favorable interface matching and exhibit much lower reflection over part of the edge-state spectrum. Nevertheless, their transmission is not uniformly reflectionless: above the edge-state minigap, the reflection increases appreciably for both geometries, becoming particularly strong for the \(60^\circ\) bend. All comparisons are performed within the approximately linear portions of the edge-state dispersion to minimize the influence of reduced group velocity on the measured reflection \cite{hughes2005extrinsic,povinelli2004effect}. Thus, the low reflection observed at selected bending angles is strongly geometry- and frequency-dependent rather than universally protected.

\clearpage
\bibliography{references}

\end{document}